\documentstyle[prb,aps,psfig,epsfig]{revtex}
\newcommand{\rf}[1]{~(\ref{#1})}

\begin{document}
\twocolumn[\hsize\textwidth\columnwidth\hsize\csname  
@twocolumnfalse\endcsname

\author{A. Perali$^{1,2}$, P. Pieri$^{1}$, 
G. C. Strinati$^{1}$, and C. Castellani$^{2}$}

\address{$^{1}$
Dipartimento di Matematica e Fisica,
Sezione INFM, Universit\`a di Camerino,\\
Via Madonna delle Carceri, I-62032 - Camerino, Italy\\
$^{2}$ 
Dipartimento di Fisica,
Sezione INFM, Universit\`a di Roma ``La Sapienza'',\\
P.le A. Moro, 2, I-00185 - Roma, Italy}

\date{\today}

\title{Pseudogap and spectral function 
from superconducting fluctuations to the 
bosonic limit}

\maketitle

\begin{abstract}
The crossover from weak to strong coupling
for a three dimensional continuum model
of fermions interacting via an attractive contact potential is studied above
the superconducting critical temperature $T_c$.
The pair-fluctuation propagator,
the one-loop self-energy, and the spectral function are investigated
in a {\em systematic way\/}
from the superconducting fluctuation regime (weak coupling) to the bosonic
regime (strong coupling). 
Analytic and numerical results are reported.
In the strong-coupling regime, where the pair fluctuation propagator
has bosonic character,  two quite different peaks appear in the spectral function
at a given wave vector, a broad one at negative frequencies and a narrow one at
positive frequencies. The broad peak of the spectral function 
at negative frequencies is
asymmetric about its maximum, with its spectral weight 
decreasing by increasing coupling and temperature. 
In this regime, two crossover temperatures $T^*_1$
(at which the two peaks in the spectral function merge in one peak)
and $T^*_0$ (at which the maximum of the lower peak crosses zero 
frequency) can be identified, with $T_c\ll T^*_0<T^*_1$.
By decreasing coupling, the two-peak structure evolves smoothly.
In the weak-coupling regime, where the fluctuation propagator has 
diffusive Ginzburg-Landau character, the overall line-shape of
the spectral function is more symmetric and  
the two crossover temperatures approach $T_c$.
The systematic analysis of the spectral function 
identifies specific features which allow one to distinguish by ARPES whether 
a system is in the weak- or strong-coupling regime. 
Connection of the results of our analysis with the phenomenology of
cuprate superconductors is also attempted and rests on the recently introduced 
{\em two-gap model\/}, according to which a crossover from weak to
strong coupling is realized when moving in the Brillouin zone
away from the nodal points toward the M points where the d-wave gap 
acquires its maximum value.
\end{abstract}
]

\section{Introduction}
\label{intro}

High-$T_c$ cuprate superconductors are characterized by doping- and
temperature-dependent anomalous properties in the metallic and
superconducting phases. At low doping (i.e., in the underdoped region of the
phase diagram), the cuprates display a {\em pseudogap} 
in the single-particle excitation spectra and in the spin susceptibility,
above the superconducting critical
temperature $T_c$ and below a crossover temperature $T^*$. The temperature
$T^*$ decreases with increasing doping and merges eventually
to $T_c$ at (or slightly above) optimum doping \cite{timstatt}.
The pseudogap phase of underdoped cuprates is best characterized by angle
resolved photoemission spectroscopy (ARPES) \cite{Ding,Norman,Campuzano}
and by tunneling experiments \cite{Renner,Miya}, which probe the 
single-particle excitation spectra directly. The pseudogap opening
below $T^*$ corresponds to a suppression of the low-frequency
differential conductance (which is connected to the density of states) 
measured by tunneling, and to a leading-edge shift of the spectral
intensity (which is connected to the spectral function
via the Fermi distribution and a dipole matrix element) measured by ARPES.
As clearly shown by ARPES, the pseudogap
is tied to the Fermi surface and its two-dimensional
wave-vector dependence is strongly anisotropic, resembling
a $d_{x^2-y^2}$ harmonic. Both ARPES and tunneling experiments
suggest that the pseudogap evolves smoothly into the superconducting 
gap as the temperature is lowered from $T^*$ to $T_c$.

The $d$-wave-like wave-vector dependence of the pseudogap, 
its continuous evolution into the superconducting gap below $T_c$, 
and its tying to the Fermi surface suggest
that the pseudogap phase could be a {\em precursor\/} of the superconducting
phase (at least in $Bi$-based compounds for which a detailed
ARPES analysis of the pseudogap is available). According to this
interpretation, the crossover temperature $T^*$ 
acquires the meaning of the temperature at which
fluctuating pairs start forming without coherence, the latter being not yet 
established owing to large fluctuations of the superconducting order 
parameter. 
Upon lowering the temperature, the coherence between pairs 
is eventually established and superconductivity appears. 
The occurrence of large superconducting pair
fluctuations in cuprates is related to the quasi-bidimensionality 
induced by the layered structure, 
as well as to the short coherence length $\xi_0$ of the 
superconducting pairs, of the order of few lattice spacing
(typically, $\xi_0 \sim 10 - 20 \AA$).

Within this scheme, the phase diagram of cuprates is interpreted in terms of a
{\em crossover\/} from Bose-Einstein (BE) condensation of preformed pairs to
BCS superconductivity, as the doping is varied
\cite{Pist94,Pist96,Andrenacci,Rand89,Levin97,Levin98,Maly,Dagotto,Yamada}.
Heavily underdoped cuprates are accordingly 
considered as superconductors in a
strong-coupling (BE) regime with $T^*\gg T_c$; optimally doped
and overdoped cuprates are instead more conventional superconductors in an
intermediate- or weak-coupling (BCS) regime with $T^* \simeq T_c$.
The evolution from strong- to weak-coupling superconductivity
as the doping is increased
is further supported by low-temperature 
ARPES and tunneling measurements in $Bi$-based compounds
of the maximum superconducting
gap $\Delta_0$ (i.e., the gap at the $M$ points of the Brillouin zone), 
whereby $\Delta_0$ decreases as the doping is increased, with
$\Delta_0 \simeq 60-70$ meV in underdoped cuprates and
$\Delta_0 \simeq 20-30$ meV in optimally 
and overdoped cuprates \cite{Renner,Miya}.
Moreover, in underdoped $Bi$-based cuprates $\Delta_0$ 
is larger than the band-width along the $M-Y(X)$ 
directions, suggesting that at least near the $M$ points
states with bosonic character can be formed.
Recent high-resolution ARPES measurements on $La_{2-x}Sr_xCuO_4$
at $T=11 K$ also indicate that the superconducting gap near the 
$M$ points increases as the doping is decreased, and smoothly evolves
into the normal phase pseudogap at very low doping \cite{Ino}.
The doping dependence of the gap is thus likely to be an universal
feature of cuprate superconductors. 

Recent high-resolution ARPES experiments in Bi2212 further
suggest that a crossover from weak
to strong coupling can even be found {\em along\/} the Fermi surface
(FS) at {\em fixed doping\/} \cite{Valla}. Fermionic states near the 
nodal ($N$) points of the FS
(namely, the points along the $\Gamma-Y(X)$ directions where
the band crosses the Fermi level
and the $d$-wave gap vanishes) appear to be weakly coupled, while
states near the $M$ points are strongly coupled and
could display bosonic character. Moving from $N$ toward $M$ 
points along the FS, a {\em continuous\/} crossover from weakly to strongly 
coupled states should accordingly be observed.
In agreement with the expectation that increasing the coupling should 
cause an increase of the width of the spectral peaks,
explicit support to the {\em wave-vector induced crossover\/} 
along the FS is obtained, for instance, from
Fig.2 of Ref.\onlinecite{Valla}. The ARPES spectral intensities
for an optimally doped Bi2212 sample reported in that figure
show, in fact, that the width of the quasi-particle
peak in the normal phase (as a function of both wave vector and frequency) 
increases along the FS, as one moves from the $N$ toward the $M$ point.
In particular, near the $M$ points the frequency distribution of the spectral 
intensity is broad and flat without any observable peak, while
a broad peak feature is present in the wave-vector distribution.
In addition, in Bi2212 at optimum doping
the band dispersion near the $M$ points along the $M-Y(X)$ direction
is rather narrow (of the order of $50$ $meV$), while the band dispersion 
along the $\Gamma-Y(X)$
directions is considerably larger (of the order of $400$ $meV$).
For all cuprates for which ARPES measurements are available,
the Fermi velocity $v_F$ is also anisotropic along the Fermi surface, with
$v_F(N)/v_F(M)\simeq 3$.\cite{Camp1}
As a consequence, fermionic states near the $M$ points are locally
associated with a small Fermi velocity 
and strong coupling ({\em hot fermions\/}); 
while fermionic states near the $N$ points are locally associated with a large
Fermi velocity and weak coupling ({\em cold fermions\/}).
To account explicitly for the different properties about the
$M$ and $N$ points, a {\em two-gap model\/} 
has been recently proposed \cite{twogap}.

In the present paper, we investigate the evolution of the spectral function 
from the weak- to strong- coupling regimes in a {\em systematic way\/},
to compare with the evolution of the spectral function in cuprates by
varying doping and wave vector. More specifically, we aim to account 
for the character of the fermionic states
near the $M$ points (where bosonic states can be formed upon reducing
the doping due to the {\em hot\/} character of these states) 
and to follow the wave-vector induced crossover 
along the Fermi surface. The local character of the fermionic states in 
wave-vector space further enables us to use a simple isotropic attraction
between electrons, which gives rise in the superconducting state to a gap
with $s$-wave symmetry. 

Two different (albeit related) kinds of approaches for the pseudogap state can be 
identified within the pairing scenario.
On the one hand, owing to the short coherence length
and the large value of the superconducting gap about the $M$ points
\cite{Pist94,Pist96,Andrenacci,Rand89,Levin97,Levin98,Dagotto,Yamada},
the superconducting phase of underdoped cuprates
is interpreted as intermediate
between a BCS state with extended pairs
and a Bose-Einstein condensate with preformed (local) pairs.
Within this view, due to strong- or intermediate-coupling effects, pairing correlations
survive well above $T_c$ and determine a pseudogap opening when
coupled to the fermions.
On the other hand, 
the second approach emphasizes the relevance of phase fluctuations
of the superconducting order parameter, owing to the
low value of the plasma frequency and the quasi-bidimensionality 
of the cuprates \cite{Emery,Sharapov}.
Within this view, the amplitude of the local order parameter is established
at $T^*$, even though phase coherence and hence long-range superconductivity
occurs at the lower temperature $T_c$.

The approach we follow in this paper belongs to the first group of the
pairing scenario. Specifically,
we investigate {\em in a systematic way\/}
the role played by pair fluctuations in
the pseudogap opening, following the BCS to Bose-Einstein crossover from
weak to strong coupling. To this end, 
we introduce a simplified microscopic model representing a 3D continuum
of fermions mutually interacting via an attractive contact potential,
which can be parametrized in terms of the scattering length.
This 3D model allows us to considerably simplify the numerical calculations as 
well to obtain analytic results (at least in some limits),
yet preserving the qualitative features obtained for more realistic
models, like the two-dimensional negative-U Hubbard model\cite{Micnas95,Singer,Vilk,Kyung}.

We examine initially the two-particle propagator in the particle-particle channel, 
and evaluate the pair-fluctuation propagator $\Gamma ({\bf k}, \omega)$
as a function of wave vector ${\bf k}$ and frequency $\omega$.
We further analyze the {\em single-particle\/} propagator,
and evaluate the self-energy $\Sigma ({\bf k},\omega)$ and
the spectral function $A({\bf k},\omega)$ within the
non-self-consistent t-matrix approximation. 
In the strong- and weak-coupling regimes, we discuss {\em analytic\/} 
forms for the self-energy, 
and comment on the main differences in the line shape
of $A({\bf k},\omega)$ between the two regimes. 
The spectral weight of the incoherent peak
that appears in $A({\bf k},\omega)$
and the temperature dependence of the chemical potential 
are also discussed. In the intermediate (crossover) region (where
analytic calculations are not feasible) only numerical results are presented.
Our findings of different characteristic features occurring for 
$A({\bf k},\omega)$ in different coupling regimes are then organized in 
a systematic way,
and a criterion to distinguish by ARPES experiments 
whether an interacting fermion system is in the strong-, intermediate-, or  
weak-coupling regime is discussed.
In the strong-coupling regime, we find it is appropriate to introduce
{\em two different crossover temperatures\/}
($T^*_1$ and $T^*_0$) to describe the peculiar evolution of
the spectral function for increasing temperature. We also show how
these two temperatures merge to a single crossover temperature
($T^*$) as the coupling is decreased.
A detailed comparison of ARPES experiments with our systematic analysis of
the spectral function in different coupling regimes is 
eventually attempted.
Although it might at first appear that our model could not be
directly applied for comparison with ARPES experiments in cuprates, 
this comparison is attempted by invoking
the wave-vector-induced crossover mentioned above.
Finally, we provide a heuristic method to rationalize the 
gross feautures of the evolution of the
spectral function in terms of the BCS spectral function, whereby the BCS gap
at $T=0$ is replaced by the value of the pseudogap at 
(or above) $T_c$ and the
Fermi energy is replaced by the renormalized chemical potential.
Even though some parts of our analysis and results have been already presented (albeit for 
different models and/or with different methods) 
in previous work\cite{Micnas95,Levin98,Maly,Yamada},
our approach should be regarded as more systematic and complete than others. 
 
The plan of the paper is as follows. In Section II we introduce the
microscopic model and discuss the relevant equations for the spectral
function and related quantities. In Sections III
and IV we report the results for two- and single-particle properties,
respectively, discussing in a {\em systematic way\/} the evolution of the 
pair-fluctuation propagator and of the spectral function from the
superconducting fluctuations regime (weak coupling) to
the bosonic limit (strong coupling).
In Section V we present a detailed comparison of our results 
with ARPES experiments. Section VI gives our conclusions.
 
\section{Relevant equations for the spectral function
and related quantities}
\label{sec1}

In this Section, we set up the relevant equations to follow
the evolution of the single-particle spectral function and the
two-particle fluctuation propagator from weak to strong coupling.
To this end, we consider a system of fermions 
embedded in a three-dimensional  continuum
and mutually interacting via an effective short-range attractive potential 
$v_0\delta ({\bf r}-{\bf r^{\prime}})$ of strenght $v_0$, where $v_0$ is a negative 
constant. 
For the 3D continuum model we are allowed to take the limit of a strictly short-range interaction
$(\sim \delta ({\bf r}-{\bf r^{\prime}}))$, thus relating to
the fermionic scattering length $a_F$ by a suitable regularization procedure.
Knowledge of the detailed form of the fermionic interaction
is, in fact, not required for studying the main features of the 
evolution from weak to strong coupling.
The many-body diagrammatic structure for the single- and two-particle
Green's functions gets in this way considerably simplified,
while preserving the physical effects of pseudogap opening \cite{nota2d3d}. 
(For a detailed discussion  of this model, see Ref.\onlinecite{Pieri}.)

For the attractive fermionic interaction of interest, the 
scattering length $a_F$ changes from being negative (when
the two-body problem fails to support a bound state) to being positive
(when the bound state is eventually supported by increasing the
interaction strength), and diverges when the coupling strength
suffices for the bound state to appear.
The dimensionless parameter $k_Fa_F$ (where $k_F$ is the Fermi wave vector)
thus locates the side of the crossover one is examining 
and how close to the crossover region one is.
Specifically, $k_Fa_F$ is small and negative in the weak-coupling 
regime, diverges in the intermediate (crossover)
regime, and eventually becomes small
and positive in the strong-coupling regime.
For this reason, driving the crossover by varying $k_F$ while
keeping $a_F$ fixed requires one to change discontinuosly
the sign of $a_F$ at the value $(k_Fa_F)^{-1}=0$.

The diagrammatic scheme we consider is based on the 
non-self-consistent t-matrix approximation, constructed with
``bare'' single-particle Green's functions (with the inclusion, however, of the
dressed chemical potential and of an additional constant energy shift (to be discussed below)
which is relevant to the symmetry of the spectral function).
This choice embodies the physics of the pseudogap state,
because in the weak-coupling regime it describes the
Ginzburg-Landau superconducting fluctuations above $T_c$ \cite{Varlamov}, 
while in the strong-coupling regime it describes the formation
of noninteracting bosons (fermionic bound states)\cite{NZSR}.

The set of relevant equations  
for the two-particle Green's function in the particle-particle
channel (i.e., the pair-fluctuation propagator), 
the single-particle Green's function, and the self-energy 
is the following:

\begin{eqnarray}
\Sigma ({\bf k},\omega_n)=-T\sum_{\nu}\int \frac{d^3q}{(2\pi)^3}
\, \Gamma^{(0)}({\bf q},\Omega_\nu)\nonumber\\
\times G^{(0)}({\bf q}-{\bf k},\Omega_\nu-\omega_n),\\
\Gamma^{(0)^{-1}}({\bf q},\Omega_\nu)=-\frac{m}{4\pi a_F}-
\int \frac{d^3k}{(2\pi)^3}\nonumber\\
\times \left[ T\sum_{n} G^{(0)}({\bf k},\omega_n)
G^{(0)}({\bf q}-{\bf k},\Omega_\nu-\omega_n)-\frac{m}{k^2}
\right],\\
G^{-1}({\bf k},\omega_n)=G^{(0)^{-1}}({\bf k},\omega_n)-
(\Sigma ({\bf k},\omega_n)-\Sigma_0),\\
n=2T\sum_{n} e^{i0^{+}\omega_n}\int \frac{d^3k}{(2\pi)^3}
\, G({\bf k},\omega_n).
\end{eqnarray}
Here, $G^{(0)}({\bf k},\omega_n)$ is the ``bare'' fermion propagator
given by $G^{(0)^{-1}}({\bf k},\omega_n)=i\omega_n-\xi ({\bf k})$
($\xi ({\bf k})={\bf k}^2/(2m)-\mu'$ being the 
free-particle dispersion measured with respect to the renormalized chemical 
potential $\mu'=\mu-\Sigma_0$, where $\mu$ is the physical chemical potential
and $\Sigma_0$ the constant self-energy shift mentioned above), $m$ is the 
free-fermion mass, and $\omega_n=\pi T(2n+1)$
($n$ integer) and $\Omega_\nu=2\pi T\nu$ ($\nu$ integer)
are, respectively, fermionic and bosonic 
Matsubara frequencies at temperature $T$.
The chemical potential is eliminated in favor of the 
density $n$ via Eq. (4). The constant self-energy shift 
$\Sigma_0$ is given by ${\rm Re}\Sigma(k=k_{\mu'},\omega=0,T\sim T^*)$ where
$k_{\mu'}=\sqrt{2m \mu'}$. 
It turns out that this shift is non-negligible only 
in the weak- to intermediate-coupling regime, where it is almost temperature
independent from $T$ to $T^*$. 
[In this coupling regime, $T^*=T^*_0=T^*_1$ is the temperature at which the pseudogap disappears.]
In practice, we will take $\Sigma_0= {\rm Re} \Sigma(k=k_{\mu'},\omega=0,T=T_c)$
in the weak- to intermediate-coupling regime, while we shall neglect 
$\Sigma_0$ altogether in the intermediate- to strong-coupling regime.
Inclusion of this self-energy shift amounts to 
a partial self-consistency dressing of the single-particle Green's functions
$G^{(0)}$, and corresponds to a complete description of the 
high-temperature region.
 
After analytic continuation to the real frequency axis \cite{notagamma},
the imaginary part of the retarded self-energy can be written at this order of 
approximation in the form:

\begin{eqnarray}
\label{imself}
\mbox{Im} \Sigma ({\bf k},\omega)&=&-\int \frac{d^3q}{(2\pi)^3}
\left [ b(\omega+\xi ({\bf q}-{\bf k}))
+f(\xi ({\bf q}-{\bf k})) \right ] \nonumber
\\
&\times& \mbox{Im}  \Gamma^{(0)}({\bf q},\omega + \xi ({\bf q}-{\bf k}))
\end{eqnarray}
where $f(x)=1/(e^{\beta x}+1)$ is the Fermi distribution
and $b(x)=1/(e^{\beta x}-1)$ is the Bose distribution, 
with $\beta= 1/T$.
Hereafter, analytic continuations are meant to produce 
retarded (R) functions.
Once the imaginary part of the self-energy is evaluated as above, 
its real part $\mbox{Re} \Sigma ({\bf k},\omega)$ is obtained 
via a Kramers-Kronig transform.
The real-frequency formulation (5) allows for high accuracy of the numerical
calculations, and avoids the problems of dealing numerically with
analytic continuation from the imaginary frequency axis.

The spectral function $A({\bf k}, \omega)$ for the single-particle fermionic 
excitations of interest is obtained from the imaginary part of the
retarded Green's function
$G^R ({\bf k},\omega)$ via the relation
\begin{equation}
\label{imgr}
A({\bf k}, \omega)=
-\frac{1}{\pi} \,\mbox{Im}  G^R ({\bf k},\omega).
\end{equation}
In terms of the real and imaginary parts of the self-energy, the spectral
function $A({\bf k}, \omega)$ has the form: 

\begin{equation}
\label{akw}
A({\bf k}, \omega)=
\frac{-\mbox{Im} \Sigma ({\bf k},\omega)/\pi}
{(\omega - \xi ({\bf k})-\mbox{Re} \Sigma ({\bf k},\omega) + \Sigma_0 )^2
+(\mbox{Im} \Sigma ({\bf k},\omega))^2} \quad .
\end{equation}

The set of equations (1)-(5), together with the definition of the
spectral function (7) and the prescription for analytic continuation
to real frequencies, allows us to study in 
a {\em systematic way\/} with limited computational effort
two- and single-particle properties over a wide range
of parameters (namely, coupling, density, and temperature), 
following the crossover from weak to strong coupling. 

A few additional comments are in order at this point about the choice (1)
of the self-energy. In the weak-coupling regime and at high enough temperature,
the expression (1) represents the leading term of a low-density 
expansion for a Fermi system even when the interaction 
is attractive\cite{Pieri}. 
Upon approaching $T_c$, the expression (1) can alternatively be interpreted
as representing the coupling of a bare fermion with pairing fluctuations.
In the strong-coupling regime, the expression (1) represents instead a
``free'' boson coupled to a bare fermion,
and is known to produce a shadow-band structure in the spectral function
at negative frequencies. We note that, to be consistent with a low-density approach, 
the ``free'' boson in Eq.(1) should be dressed with suitable self-energy 
corrections for composite bosons (as discussed in Ref.\onlinecite{Pieri}), at least in the
not-too-extreme strong-coupling regime where the residual interaction
between the composite bosons remains active.
We defer this study to a future work, and regard the present
approach with a ``free'' boson only as a preliminary but essential
step to establish the qualitative behavior of the spectral function
in a systematic way, for all coupling regimes and over a wide temperature 
range. 

Finally, we recall that inclusion of full self-consistency
in the single-particle Green's functions entering Eq.(1) can be
safely dismissed (at least in the weak- and strong-coupling regimes,
where the use of expression (1) can be justified to start with),
as discussed in Ref.\onlinecite{Pieri}. 
In contrast to the non-self-consistent t-matrix approximation used in Eqs. (2-4), 
the self-consistent t-matrix uses full self-consistent Green's functions 
but does not include vertex corrections in the self-energy.
As remarked in Ref.\onlinecite{Kyung}, the different levels of approximation for 
vertexes and single-particle Green's functions may then lead to 
unphysical results for the pseudogap in the spectral function, in a similar way to
what happens for the 2D repulsive Hubbard model \cite{Moukouri}.
The important point to be emphasized is that, provided $G^{(0)}$ in Eqs.(1) and (2)
contains the dressed chemical potential obtained from Eqs.(3) and (4), this
set of equations interpolates smoothly between weak- and strong-coupling limits and provides a 
reasonable description of both limits.
Moreover, we emphasize again that our interest in the (three-dimensional) continuum 
model originates essentially from obtaining analytic results in the continuum 
case, from which the main features of the spectral function can be readily  
extracted. A comparison between our systematic results for the continuum
model and those available for the lattice model will be made in 
Section VI, with the outcome that the main qualitative features remain the
same in the two models.

\section{Pairing fluctuations from weak to strong coupling}
\label{sec2}

In this Section, a {\em systematic study\/} of
the crossover from the superconducting fluctuation (weak-coupling) regime 
to the bosonic (strong-coupling) regime for the
pair-fluctuation propagator is reported via both
analytical and numerical calculations,
for a wide temperature range above $T_c$. This study is preliminary to the
discussion of the spectral function via the self-energy (1),  
presented in the next Section.

The pair-fluctuation propagator $\Gamma^{(0)}({\bf q},\Omega_\nu)$
within the non-self-consistent t-matrix approximation is given by Eq.(2).
We shall examine, in particular, its wave-vector and frequency 
dependence for all coupling regimes.

Being interested in {\em normal-phase\/} properties of the fermionic system, 
knowledge of the superconducting critical 
temperature $T_c$ is required at the outset to insure that $T\ge T_c$.
To identify the critical temperature $T_c$, we rely on
the condition that the fluctuation propagator has a pole 
at $T_c$ for vanishing wave vector and frequency, namely, 
$\Gamma^{(0)^{-1}}({\bf q}=0,\Omega_\nu=0; \mu,T=T_c)=0$.
This condition (known as the Thouless criterion for the 
superconducting instability) taken alone is equivalent to the BCS
equation for the critical temperature in the weak-coupling limit.
In addition, when coupled to the density equation (4) to fix the 
chemical potential, it yields the value of the 
Bose-Einstein condensation temperature in the strong-coupling limit. 

The Thouless criterion provides a temperature-dependent {\em critical\/} value for the 
chemical potential $\mu_c(T_c)=\mu (n,T_c)$, with $\mu (n,T)$ obtained from
the density equation. 
(Recall once more that, for a proper description of the crossover from weak to
strong coupling, it is {\em essential\/} to let the chemical potential
adjust itself with coupling at given density.)
In Fig.1
the critical chemical potential
and the normal-phase chemical potential are reported 
for different values of $(k_Fa_F)^{-1}$ 
in the strong- to intermediate-coupling regime.
We have verified that in the high-temperature limit 
the chemical potential $\mu (n,T)$
tends to its classical value
$\mu (n,T)=1.5T\ln (1.898 \, T_{BE}/T)$ for all densities ($T_{BE}=3.31 n_B^{2/3}/m_B$ being the 
Bose-Einstein condensation temperature).\cite{SVR}
\begin{figure}
\centerline{\psfig{figure=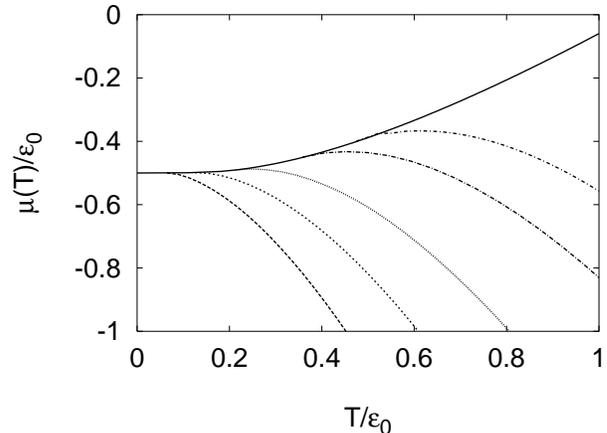,width=6cm,angle=-90}}
\caption{Critical chemical potential $\mu_c(T_c)$
(full line)
and normal-phase chemical potential $\mu(n,T)$ for different values of the 
density expressed in terms of $(k_Fa_F)^{-1}$ (broken lines), as functions of 
temperature. The values of $(k_Fa_F)^{-1}$ are 1.33,0.96,0.76,0.60, and 0.52 from
bottom to top.}
\label{f1}
\end{figure}

Note that in the strong-coupling regime, i.e., at low density and for 
$\beta|\mu|\gg 1$, $\mu$ approaches the value 
$-\epsilon_0/2$, where $\epsilon_0=(ma_F^2)^{-1}$ is the binding energy
of the associated two-body problem.
In the intermediate-coupling regime, $\mu$ initially increases as 
the temperature is increased, reaches a maximum, and eventually decreases, 
tending to the classical behavior. 
The temperature where the maximum is located 
turns out to be smaller than the binding energy ${\epsilon_0}$,
and does not relate to specific changes of single-particle properties
in the strong- or intermediate-coupling regimes. 
In the weak-coupling regime, on the other hand, this temperature 
turns out to coincide with the crossover temperature $T^*$ where pair 
fluctuations become manifest and a pseudogap opens.

In the weak-coupling regime, the presence of the maximum in $\mu(T)$
is connected with the fermionic degrees of freedom,
and in particular with the opening of a pseudogap at the Fermi surface
which tends to depress the chemical potential upon lowering the temperature 
(in a similar fashion to what happens when a real gap opens 
in a BCS superconductor below $T_c$). 
The presence of a maximum of $\mu(T)$ 
(with the ensuing non monotonic behavior of $\mu(T)$)
is clearly observed in Monte Carlo
simulations of the 2D attractive ($s$-wave) Hubbard model in the
intermediate-coupling regime (cf. Fig.6 of Ref.\onlinecite{Singer}),
while the presence of the maximum is somewhat debated
when the self-consistent t-matrix approximation is 
used \cite{Micnas95,Gooding}.
In the strong-coupling regime, the fermionic degrees of freedom are 
exponentially suppressed according to $f(\xi)\sim \exp (-\beta | \mu |)$ and 
the above
maximum is progressively shifted toward zero temperature 
for increasing $\beta| \mu |$, thus recovering in the extreme strong-coupling limit
the behavior of
a free Bose gas via the relation $2\mu=-\epsilon_0 +\mu_B$\cite{Hauss94,haus,Pist96}. We have 
verified that, in the strong- and intermediate-coupling regimes, the temperature at which
$\mu (T)$ reaches its maximum does not relate with the
temperature at which the pseudogap opens.  

The equation for the density, together with the condition for the
critical chemical potential $\mu (T)=\mu_c(T)$, yields
the value of the critical temperature $T_c$
for the superconducting instability. 
The critical temperature $T_c$ 
and the BCS mean-field critical temperature $T_{BCS}$
(the latter obtained formally from the same equations 
defining $T_c$ but with the {\em bare\/}
single-particle Green's function $G^{(0)}$ replacing $G$ in Eq.(4))
are reported in Fig.2 as functions of the parameter $(k_Fa_F)^{-1}$
(recall that in the weak-coupling limit $T_{BCS}=1.67 \epsilon_F
\exp({\pi/2k_Fa_F})$ ($a_F<0$)).
(Both temperatures have been conveniently normalized 
to the Bose-Einstein condensation temperature 
$T_{BE}$ evaluated at the same density. Note that in the 3D continuum 
model $T_{BE}$ is of the same order of $\epsilon_F$.)
The results of Fig.2 can be compared with the calculation of $T_c$ within
the non-self-consistent t-matrix approximation reported in
Ref.\onlinecite{Hauss94}.
We mention that the inclusion of the self-energy shift $\Sigma_0$ in the weak- to 
intermediate-coupling regime adopted in the present paper, slightly increases 
the value of the critical temperature with respect to the results of
Ref.\onlinecite{Hauss94}.
\begin{figure}
\centerline{\psfig{figure=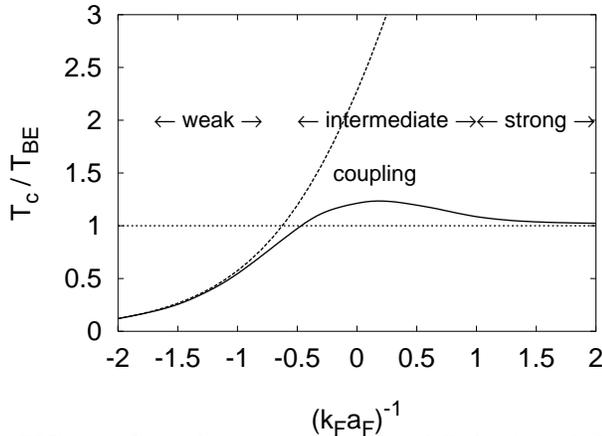,width=6cm,angle=-90}}
\caption{Critical temperature $T_c$ (full line) and BCS mean-field critical
temperature $T_{BCS}$ (broken line) as functions of $(k_Fa_F)^{-1}$ 
(both temperatures are normalized to the
Bose-Einstein condensation temperature $T_{BE}$ evaluated at the
same density).}
\label{tctbe}
\end{figure}
We have verified that, in the strong- and intermediate-coupling regimes,
the mean-field temperature $T_{BCS}$ about 
coincides with the temperature at which the bare (fermionic)
contribution $n_0$ to the total density $n$ equals the
(bosonic) contribution $\delta n$ due to interaction effects, i.e.,
$n_0(T=T_{BCS})\simeq \delta n(T=T_{BCS}) \simeq n/2$.
This result permits us to identify $T_{BCS}$ as
the {\em crossover temperature where preformed pairs 
start to form\/}. [The connection between $T_{{\rm BCS}}$ and the characteristic 
crossover temperature(s) of the spectral function will be made in Section IV.]  
Note from Fig.2 that for
$(k_Fa_F)^{-1}\lesssim -1$ the critical temperature approaches
the BCS mean-field critical temperature, indicating that the
fermionic system is in the weak-coupling regime.
For $(k_Fa_F)^{-1}\gtrsim 1$ 
the critical temperature is instead close to
the Bose-Einstein condensation temperature, 
indicating that in the strong-coupling regime the
fermionic system is equivalent to
a system of non-interacting bosons. 
The strong-coupling limit is thus effectively
reached for not too large values of the parameter 
$(k_Fa_F)^{-1}$. \cite{notakfxi}

In Fig.3 we report for convenience the relation between $k_F\xi_{pair}$ and
$(k_Fa_F)^{-1}$ as obtained from the analytic solution of Ref.\onlinecite{Marini}, 
$\xi_{pair}$ being the average pair size at $T=0$.
This plot is especially useful to compare our results for the two- and
single-particle properties (which are expressed in terms of $(k_Fa_F)^{-1}$)
with the phenomenology of cuprates, for which some estimates
of the parameter $k_F\xi_{pair}$ in 
different doping regimes are available \cite{Pist94}.
\begin{figure}
\centerline{\psfig{figure=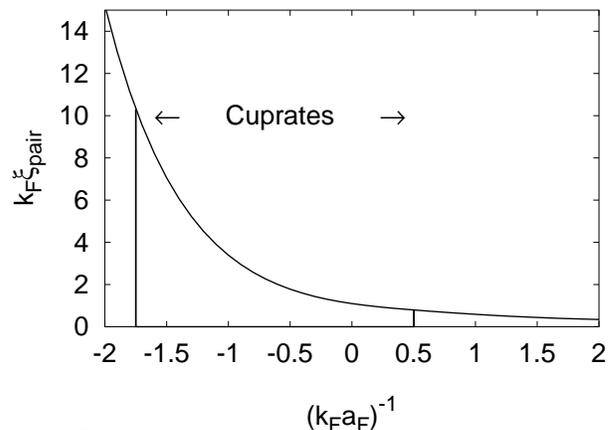,width=6cm,angle=-90}}
\caption{Phenomenological parameter $k_F\xi_{pair}$
as a function of the coupling parameter $(k_Fa_F)^{-1}$, from weak to strong
coupling.}
\label{kfxi}
\end{figure}
Specifically, at optimum doping the parameter 
$k_F\xi_{pair}$ takes roughly values between 6 and 10, and
its value decreases for decreasing doping (mainly because the
superconducting gap at $T=0$ increases and the Fermi
energy decreases approaching the insulating phase).
We may reasonably consider $k_F\xi_{pair}\approx 1$ as a lower-bound 
value for underdoped cuprates, especially if we consider it as a local quantity
about the $M$ points.
Accordingly, the coupling parameter $(k_Fa_F)^{-1}$
in the optimum and underdoped regimes for cuprates lies approximately in the
range $-1.7\,\lesssim\,(k_Fa_F)^{-1}\,\lesssim\,0.5$, 
as indicated in Fig.3.

Having determined the thermodynamic quantities $\mu (n,T)$ 
and $T_c$, we pass now to calculate the pair-fluctuation 
propagator (2). From a physical point of view, pairing fluctuations 
have essentially different character in the
strong- and weak-coupling regimes.
While to evaluate the self-energy numerically from Eq.(5) knowledge 
of $\Gamma^{(0)}$ is required over a wide range of wave vectors and 
frequencies, to characterize the evolution of
$\Gamma^{(0)}$  from weak to strong coupling it is sufficient to consider
the expansion of its inverse
in powers of the wave vector ${\bf q}$ and the Matsubara 
frequency $\Omega_{\nu}$:\cite{Maly,Yamada}

\begin{equation}
\label{ggen}
\Gamma^{(0)^{-1}}({\bf q},\Omega_{\nu})=
a+b| {\bf q}| ^2+d\,i\Omega_{\nu}
\end{equation}
with $d=d_1+id_2\mbox{sgn}(\Omega_{\nu})$.
Here, $(a,b,d_1,d_2)$ are real coefficients which
are coupling, density, and temperature dependent.

In Fig.~4 we report the ratio of the imaginary ($d_2$) 
to the real ($d_1$) part of the frequency coefficient in Eq.(8) at $T_c$
as a function of $(k_Fa_F)^{-1}$.
\begin{figure}
\centerline{\psfig{figure=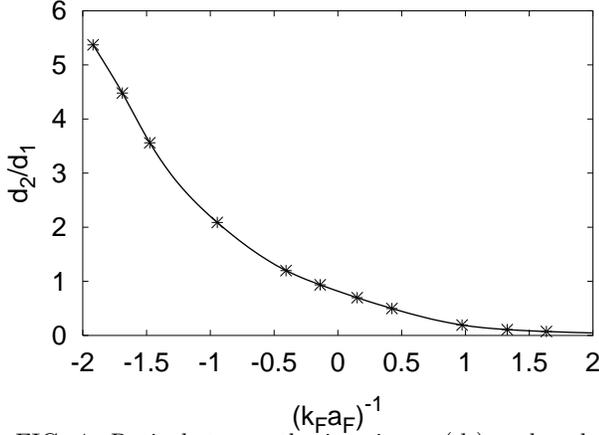,width=6cm,angle=-90}}
\caption{Ratio between the imaginary $(d_2)$ and real $(d_1)$ part
of the frequency coefficient of the inverse pair-fluctuation propagator at 
$T_c$ as a function of $(k_Fa_F)^{-1}$.}
\label{d2d1}
\end{figure}
In the strong-coupling limit, $d_1 \simeq -m^2 a_F/(8\pi)$ and 
$d_2 \simeq 0$, 
with the pair-fluctuation
propagator acquiring the polar structure of a 
bosonic Green's function. 
In the weak-coupling limit, on the other hand, 
$d_1 \sim -(T_c/E_F)^2\ll 1$ and $d_2=-N_0\pi/(8T_c)$,
where $N_0$ is the density of states (per spin component)
at the Fermi energy $E_F$,
with the pair-fluctuation propagator acquiring the
diffusive Ginzburg-Landau structure \cite{Varlamov}. 
\begin{figure}
\centerline{\psfig{figure=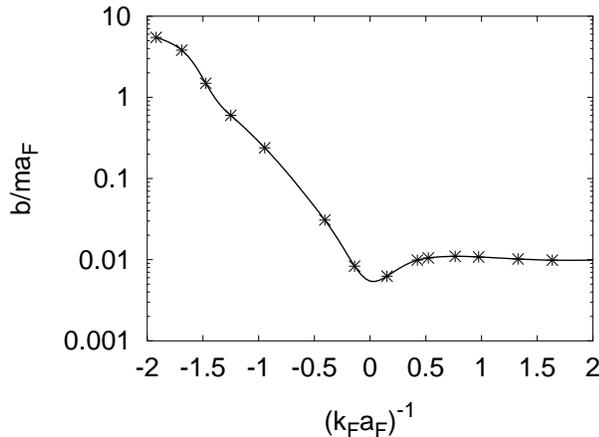,width=6cm,angle=-90}}
\caption{Coefficient $b$ of the $|{\bf q}|^2$ term in the inverse pair-fluctuation
propagator at $T=T_c$ as a function of 
$(k_Fa_F)^{-1}$.}
\label{bcoeff}
\end{figure}
When $| d_1| > | d_2|$, fluctuating pairs become propagating 
(albeit with a damping), and eventually acquire the (bosonic) character 
of undamped preformed pairs.

In Fig.5 the coefficient $b$ of the $|{\bf q}|^2$ term in Eq.(8)
is plotted as a function of $(k_Fa_F)^{-1}$.

In the strong-coupling regime (i.e., for $(k_Fa_F)^{-1}\gtrsim1$), 
$b$ coincides with its strong-coupling value
$ma_F/(32\pi)$, where $2m$ is the mass of the composite boson.
In the weak-coupling regime (i.e., for $1/k_Fa_F\lesssim -1$), 
$b= N_0 7\zeta(3)v_F^2/(48\pi^2\,T_c^2)$ 
is proportional to the square of the zero-temperature correlation  
length $\xi_0$
(where $v_F$ is the Fermi velocity and $\zeta(3)\simeq 1.202$ 
is the Riemann zeta function of argument 3),
and tends to diverge in the extreme weak-coupling limit.  

Finally, the coefficient $a$ in Eq.(8) 
provides a mass to the pair-fluctuation propagator.
In the weak-coupling limit $a=N_0\ln (T/T_c)$ vanishes at $T_c$, while in the
strong-coupling limit we can write $a=-m^2a_F\mu_B(T)/(8\pi)$ where $\mu_B(T)$
is the bosonic chemical potential, which we pass now to discuss.

According to the above analysis, we have verified that in the 
{\em strong-coupling\/} regime (when the conditions $\mu < 0$ and 
$\beta | \mu | \gg 1$ are satisfied)
the pair-fluctuation propagator evaluated numerically
acquires the polar structure of a free-boson Green's 
function\cite{haus,Pieri}: 
\begin{equation}
\label{gbos}
\Gamma^{(0)}({\bf q},\Omega_{\nu})=-\frac{8\pi /(m^2 a_F)}
{i\Omega_{\nu}-q^2/(4m)+\mu_B}
\end{equation}
with mass $2m$, a quadratic dispersion as a function of wave vector,
and bosonic chemical potential $\mu_B(T)$, which 
reduces to $2\mu(T)+\epsilon_0$ in the extreme strong-coupling limit. 
This implies that, for real frequencies,
the imaginary part of $\Gamma^{(0)}$
(which enters the calculation of the imaginary part of the
self-energy via Eq.(5)) is proportional to a delta function 
in the strong-coupling limit:

\begin{equation}
\label{imgbos}
\mbox{Im} \Gamma^{(0)}({\bf q},\omega)=\frac{8\pi^2}{m^2 a_F}\,
\delta (\omega-q^2/(4m)+\mu_B).
\label{deltag}
\end{equation}

We have verified that Eq.(\ref{imgbos}) remains approximately 
valid in the {\em intermediate-coupling\/} regime toward strong coupling (when 
$\beta |\mu| \sim 1$ and $\mu <0$), while only in the strong-coupling limit
$\mu_B(T)$ therein reduces to the chemical
potential $\mu_0 (T)$ of an ideal Bose gas,
with the characteristic temperature dependence
\begin{equation}
\label{mubt}
\mu_0 (T)=-1.22\, \frac{(T-T_{BE})^2}{T_{BE}}
\end{equation}
at low enough temperature.
Specifically, we have verified that a 
delta-function contribution to $\mbox{Im} \Gamma^{(0)}({\bf q},\omega )$
appears when $(k_Fa_F)^{-1}\ge 0$ (corresponding to $d_2/d_1\lesssim 1$ in 
Fig.4). This contribution, which is initially present for large values of 
${\bf q}$, extends progressively to smaller values of ${\bf q}$ for increasing 
coupling, reaching eventually ${\bf q}=0$ when the chemical potential becomes negative.
By further increasing the coupling, the delta-function contribution to 
 $\mbox{Im} \Gamma^{(0)}({\bf q},\omega )$  becomes increasingly prominent and 
the asymptotic expression\rf{deltag} is progressively reached.  

This delta-function contribution, associated with the formation of a bound state
with bosonic character, is responsible in the  strong-coupling limit for the opening of a 
real gap in a broad temperature range above $T_c$, as it will be shown in the next 
Section. Actually, in addition to the delta function, one also finds a finite
contribution to $\mbox{Im} \Gamma^{(0)}$ for $\omega >2 \mid \mu \mid$. When
inserted in Eq. (5), this contribution leads to an exponentially vanishing 
$\mbox{Im} \Sigma$ for $\omega > \mid \mu \mid$.

Finally, in the {\em weak-coupling\/} regime, the pair-fluctuation propagator 
recovers the Ginzburg-Landau diffusive form.
Near the critical temperature, its expression for small wave vectors 
and frequencies is accordingly given by: 

\begin{equation}
\label{glprop}
\Gamma^{(0)}({\bf q},\Omega_{\nu})
=\frac{1}{N_0\left(\varepsilon + \eta |{\bf q}|^2 +
\gamma |\Omega_{\nu} |\right)}\, .
\end{equation}
Here, $\varepsilon =\ln (T/T_c)$ is the mass term 
of the propagator,
$\eta \sim v_F^2/T_c^2$ represents the stiffness of the
superconducting fluctuations with a proportionality coefficient 
which depends on dimensionality  ($7\zeta (3)/(48\pi^2)$ being its value in 3D), 
while $\gamma = \pi/(8T_c)$ is related to the lifetime of 
the fluctuating pairs (which do not obey Bose statistics).
In this limit, $\mbox{Im} \Gamma^{(0)}({\bf q}=0,\omega )$ 
diverges like $1/\omega$
only at the critical temperature $T=T_c$; as a consequence,
the pseudogap region induced by the diffusive pair fluctuations will be 
present only in a rather {\em narrow\/} temperature range. No delta function contributes 
in this regime, but $\mbox{Im} \Gamma^{(0)}({\bf q}=0,\omega )$ 
has a broadened peak structure for small enough ${\bf q}$.

To summarize, the main effect of increasing coupling  in
$\mbox{Im} \Gamma^{(0)}({\bf q},\omega )$ 
is the appearance of a {\em peak structure\/} (delta function)
at finite frequencies, whose area grows with $(k_Fa_F)^{-1}$. 
In the strong-coupling regime, for ${\bf q}=0$ and
$T>T_c$ the real part of $\Gamma^{(0)}({\bf q},\omega )^{-1}$
vanishes at a finite frequency, corresponding to a pair resonance.
This resonance disperses as ${\bf q}^2$.
In the weak-coupling regime, the real part 
of $\Gamma^{(0)}({\bf q},\omega )^{-1}$ is small only in the 
critical region, 
and vanishes only at the critical temperature.
Increasing the coupling from weak to strong,
the frequency dependence of $\mbox{Im} \Gamma^{(0)}({\bf q}=0,\omega )$ 
evolves from being antisymmetric 
with respect to $\omega =0$ to an asymmetric structure. 
This evolution confirms 
previous results for $\Gamma^{(0)}({\bf q}=0,\omega )$ reported
in Ref.\onlinecite{Maly}.

In the next Section, we will show how the peak structures of 
$\Gamma^{(0)}({\bf q},\omega )$ affect the single-particle self-energy and 
hence the spectral function, giving rise to a (pronounced) 
suppression of the low-energy spectral weight, namely, to a pseudogap.

\section{Spectral function from weak to strong coupling}
\label{Sec4}

In this Section, we study the single-particle excitations for fermions
coupled to pair fluctuations above the critical temperature.
The spectral function $A({\bf k},\omega)$, obtained by solving the
set of equations (1)-(7), is analyzed in 
{\em a systematic way\/} as a function of coupling and temperature, thus 
following its evolution from weak to strong coupling.
In this way, characteristic features of the spectral function as a 
function of frequency and temperature will be evidenced in {\em all\/}
coupling regimes. We shall analyze separately
the cases when the chemical potential lies below the bottom of the
single-particle band ($\mu<0$) corresponding to the strong- to
intermediate-coupling regime, and when the chemical
potential lies inside the single-particle band ($\mu>0$) corresponding
to the intermediate- to weak-coupling regime.

\subsection{Strong- to intermediate-coupling regime} 

In the previous Section, we have verified that a delta function appears 
in $\mbox{Im}\Gamma^{(0)}({\bf q},\omega)$
starting from the intermediate-coupling regime when $(k_Fa_F)^{-1} >0$ 
(for temperatures such that $\beta| \mu|\gtrsim1$).
In particular, we have verified that
in the {\em strong-coupling limit\/} (where 
$\mu<0$ and $\beta |\mu| \gg 1$) 
the pair-fluctuation propagator coincides
with a free-boson Green's function with mass $2m$ (cf. Eq.(9)). 
In this limit, the imaginary part of
the pair-fluctuation propagator reduces to a delta function and the 
self-energy can be evaluated analitically.
Inserting Eq.(10) into the general expression (5) for
the imaginary part of the self-energy, the following form results:

\begin{equation}
\label{imselfan}
\mbox{Im} \Sigma ({\bf k}=0,\omega)=
\frac{-2(4m)^{3/2}}{m^2a_F}
\frac{\sqrt{\omega_{th}-\omega} \Theta (\omega_{th}-\omega)}
{e^{\beta (\omega_{th}-\omega+|\mu_B|)}-1}
\end{equation}
where $\omega_{th}=\mu-\mu_B$ is a {\em threshold\/} frequency and
$\Theta$ is the unit step function. 
[In the strong- to intermediate-coupling regime, 
when the chemical potential
is below the bottom of the free-fermion band $(\mu <0)$ 
and ${\bf k} ^2/(2m)\ll \epsilon_0$, the self-energy 
and hence the spectral function are almost independent of wave vector.
In this case, ${\bf k}=0$ can be taken as a representative value, as we did
in Eq.(13).]

Note that the frequency dependence of $\mbox{Im} \Sigma$ 
is strongly {\em asymmetric\/} about its minimum at 
$\omega \simeq \omega_{th}-|\mu_B|$.
Note also that $\mbox{Im}\Sigma$ (and hence $\mbox{Re}\Sigma$ obtained
via Kramers-Kronig transform) has a nontrivial
temperature and frequency dependence, showing strong
deviations from Fermi-liquid behavior.
In the regime where $\beta |\mu_B| << 1$ ({\em i.e., $T\simeq T_c$}), 
three different behaviors of $\mbox{Im}\Sigma (0,\omega)$
can be specifically identified
on the frequency axis: ({\em i}) For $\omega_{th}-|\mu_B|<\omega <\omega_{th}$,
$\mbox{Im}\Sigma (0,\omega) \sim -\sqrt{\omega_{th}-\omega}/\beta |\mu_B|$;
({\em ii}) For $\omega_{th}-\beta <\omega < \omega_{th}-|\mu_B|$,
$\mbox{Im}\Sigma (0,\omega) \sim -(\beta \sqrt{\omega_{th}-\omega})^{-1}$; 
({\em iii}) For $\omega < \omega_{th}-\beta$, 
$\mbox{Im}\Sigma (0,\omega) \sim -\exp (-\beta(\omega_{th}-\omega))$.
Note that, in the strong-coupling limit, the imaginary part of the 
self-energy has a square-root divergence at $\omega = \omega_{th}$
for $T=T_c$.

We have further verified numerically that strong deviations from 
Fermi-liquid behavior are present in the strong- to intermediate-coupling 
regime in a wide temperature range above $T_c$ 
(while in the weak-coupling regime non-Fermi-liquid behavior is 
found only in a narrow temperature range above $T_c$, as 
discussed in the next subsection). 

The above characteristic features of the analytic expression (13) can be
clearly identified in the numerical results 
for $\mbox{Im}\Sigma({\bf k}=0, \omega)$ reported in Fig.6
at different temperatures
(for a specific coupling). The associated real part is shown in
Fig.7, where the straight lines $\omega +\mu$ are also reported for
the same temperatures (and coupling), with increasing temperature from top to 
bottom. 
At any temperature, the intersection of a given straight line with
$\mbox{Re}\Sigma({\bf k}=0,\omega)$ locates the position of the
{\em quasi-particle peak} at $\omega >0$. 

\begin{figure}
\centerline{\psfig{figure=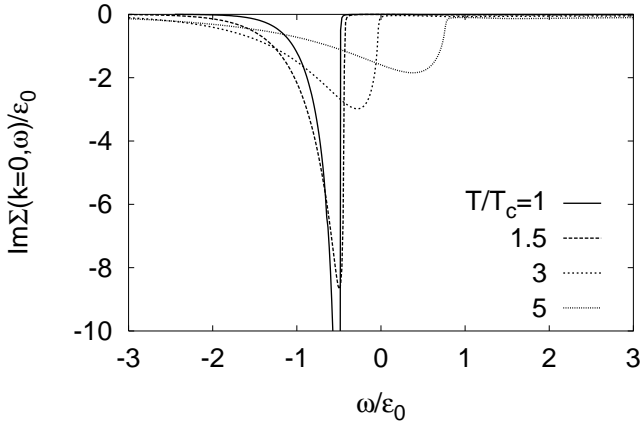,width=6cm,angle=-90}}
\caption{Imaginary part of the self-energy at ${\bf k}=0$ 
as a function of frequency (in units of $\epsilon_0$) at different 
temperatures.
In this case, $(k_Fa_F)^{-1}=0.77$ and $T_c/T_{BE}=1.14$.
(Strong- to intermediate-coupling regime.)}
\label{selfsci}
\end{figure}

\begin{figure}
\centerline{\psfig{figure=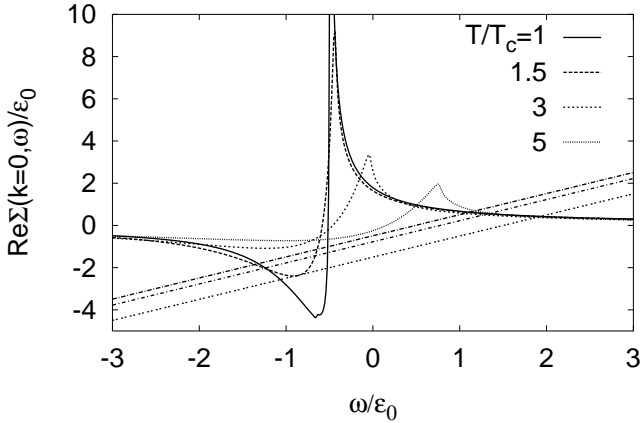,width=6cm,angle=-90}}
\caption{Real part of the self-energy at ${\bf k}=0$ 
as a function of frequency (in units of $\epsilon_0$) 
at different temperatures.
In this case, $(k_Fa_F)^{-1}=0.77$ and $T_c/T_{BE}=1.14$.
(Strong- to intermediate-coupling regime.)}
\label{selfscr}
\end{figure}

\begin{figure}
\centerline{\psfig{figure=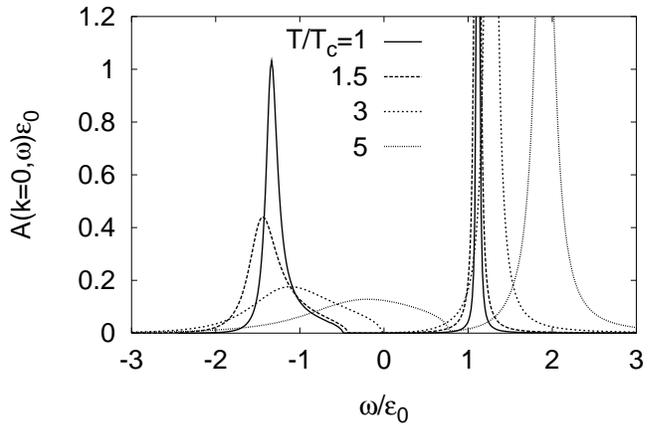,width=6cm,angle=-90}}
\caption{Spectral function at ${\bf k}=0$ 
as a function of frequency (in units of $\epsilon_0$) at different 
temperatures.
In this case, $(k_Fa_F)^{-1}=0.77$ and $T_c/T_{BE}=1.14$.
(Strong- to intermediate-coupling regime.)}
\label{akwsct}
\end{figure}

Note that for temperatures close to
$T_c$, three intersections occur, with the most left intersection giving rise to 
the {\em incoherent peak} in $A({\bf k},\omega)$ at negative frequencies
(while to the central intersection there corresponds a strong suppression of
$A({\bf k},\omega)$). 
At high enough temperatures, on the other hand, only a single intersection 
occurs. The corresponding spectral function for the same coupling
and temperatures above $T_c$ is reported in Fig.\ref{akwsct}.
The resulting spectral function has a {\em strongly asymmetric structure with 
two peaks\/}:
The one at positive frequencies is rather narrow, coherent like, 
and has a large spectral weight (namely, the area enclosed by the peak); 
The one at negative frequencies is instead broad and has
a small spectral weight. 
When the chemical potential $\mu$ is below the bottom
of the band (as in the strong-coupling limit), the peak located at
negative frequencies represents the {\em incoherent peak\/} generated by 
the interaction of the fermions with strong pair fluctuations 
(recall that, in this limit, 
the pair-fluctuation propagator has the polar structure
of a bosonic Green's function). This incoherent peak is itself asymmetric,
it becomes broader for increasing temperature,
its spectral weight is density and coupling dependent (decreasing like
$(k_Fa_F)^3$), and its position depends mainly on the 
value of the chemical potential (which in turn depends on temperature). 
By increasing temperature,
the chemical potential becomes progressively more negative
(cf. Fig.\ref{f1})
and the {\em peak position of $A({\bf k=0},\omega)$
shifts accordingly toward positive frequencies\/}.
The broadening of the incoherent peak becomes pronounced when
the temperature is of the order of the binding energy
(see, e.g., the case with $T/T_c=3$ in Fig.8).
Note also that, for increasing temperature, the two peaks in 
$A({\bf k=0},\omega)$ get broadened in an asymmetric way
(in contrast to the weak-coupling regime (see below), approaching
which the broadening of the two peaks becomes progressively more
symmetric). We have also verified that,
in the extreme bosonic limit, the spectral function
has the structure of two delta-like peaks symmetrically 
located with respect to $\omega =0$ (albeit with quite different spectral weights), 
which is generated in an asymmetric way by
the narrowing of the incoherent peak at negative frequencies
as the product $k_Fa_F$ becomes smaller and smaller.

\begin{figure}
\centerline{\psfig{figure=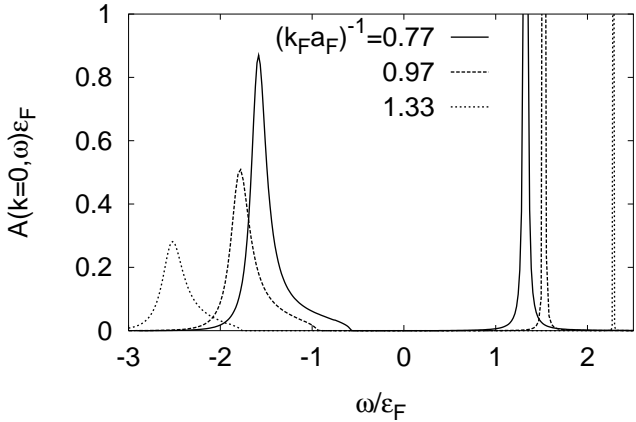,width=6cm,angle=-90}}
\caption{Spectral function at ${\bf k}=0$ 
as a function of frequency (in units of $\epsilon_F$) at $T=T_c$ for different values of 
the coupling $(k_Fa_F)^{-1}$. (Strong- to intermediate-coupling regime.)}
\label{fakwkf}
\end{figure}

In Fig.\ref{fakwkf} the spectral function at ${\bf k=0}$ is plotted
as a function of frequency for different values of the parameter 
$(k_Fa_F)^{-1}$ at $T=T_c$. Note that these curves have been expressed in units
$\epsilon_F$, instead of $\epsilon_0$, to get a more evident evolution with 
coupling.
The spectral function has two well-separated peaks, with {\em a real gap\/} opening 
at an energy of the order of the binding energy of the pairs.
By increasing the coupling, the spectral weight inside the gap
is progressively suppressed, until in the extreme strong-coupling limit the 
step function in the imaginary part of the self-energy (13) makes
the spectral weight to vanish identically in the range 
$-| \mu|+| \mu_B| < \omega < | \mu|$. 
 
Since photoemission experiments measure the intensity of photo-emitted 
electrons (that is, the spectral weight at {\em negative\/} frequencies), no 
signal would be detected if both the incoherent and coherent peaks had moved 
to positive frequencies for increasing temperature.
In this context (and in analogy with what is empirically done when interpreting
photoemission measurements), it is natural to introduce a {\em crossover 
temperature\/} $T^*_0$
at which the maximum of the lower peak 
crosses zero frequency.
Our analysis shows, however, that at $T^*_0$ the spectral
function still maintains a two-peak structure 
(cf. Fig.\ref{akwsct}), reflecting the sizable effects of the interaction 
between fermions and pair fluctuations.
We are accordingly led to introduce a {\em second crossover 
temperature\/} $T^*_1>T^*_0$, at which the upper and lower peaks
of the spectral function merge just in one peak (in the sense that the incoherent
peak is progressively absorbed by the coherent (quasi-particle) peak, even 
though the separation between the two peaks remains almost constant).
[Photoemission measurements alone, however, would not be able to identify this 
second crossover temperature $T_1^*$ since the merging of the two peaks would occur
at positive frequencies, a region which only {\em inverse} photoemission is
able to probe.]

In Fig.10 the two crossover temperatures    
$T^*_0$ and $T^*_1$ (as obtained numerically from the above definitions) 
are reported as functions of the parameter $(k_Fa_F)^{-1}$, 
both temperatures being normalized with respect to 
the critical temperature $T_c$. The BCS mean-field critical temperature 
$T_{{\rm BCS}}$ from Fig.~2 is also reported for comparison.
In the strong-coupling limit (when 
$(k_Fa_F)^{-1} \gtrsim 1$) $T^*_1 \gg T^*_0$,
$T^*_1$ being a large energy scale which, according to Fig. 10,
in the strong-coupling limit is much larger than the binding energy $\epsilon_0$.
The difference between $T^*_1$ and $T^*_0$ is reduced
by decreasing $(k_Fa_F)^{-1}$, 
but only in the intermediate-coupling regime 
(i.e., when $(k_Fa_F)^{-1} \lesssim -0.1$)
the two crossover temperatures almost coincide 
($T^*_1 \simeq T^*_0$).
In the weak-coupling regime, only a single crossover temperature 
can be identified
($T^*_1=T^*_0$), since in this regime the chemical potential is 
almost equal to the Fermi energy and the 
two peaks of the spectral function are symmetrically located about
zero frequency. Note finally that $T_0^*$ about coincides with $T_{{\rm BCS}}$ 
which was previously identified via an independent procedure. [When 
$T_0^*\simeq T_1^*$ we shall indicate both temperatures simply as $T^*$.]
\begin{figure}
\label{tcts}
\centerline{\psfig{figure=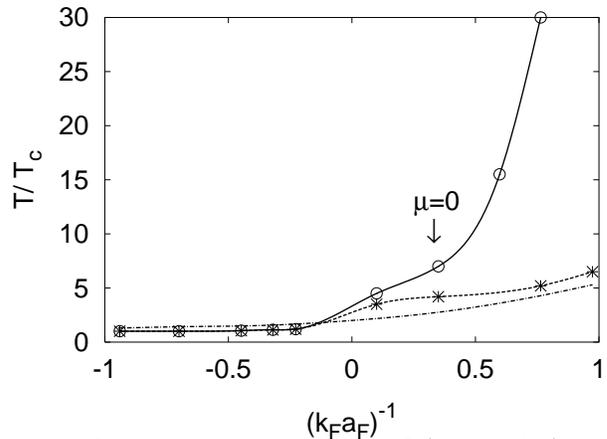,width=6cm,angle=-90}}
\caption{Crossover
temperatures $T^*_0$ (dashed line) and $T^*_1$ (full line), and BCS critical temperature
$T_{{\rm BCS}}$ (dashed-dotted line) as functions of $(k_Fa_F)^{-1}$; all temperatures are
normalized to the critical temperature $T_c$ of Fig.~2.
The value of $(k_Fa_F)^{-1}$ where the chemical potential changes
sign is indicated by an arrow.}
\end{figure}

\subsection{Intermediate- to weak-coupling regime}
In the intermediate to weak-coupling regime, it becomes essential to take 
explicit account of the constant shift $\Sigma_0$ introduced in Section II.
This shift has been identified with the value of the real part 
of the self-energy (1) taken at the wave-vector $k_{\mu'}$ (such that 
$\xi(k_{\mu'})=0$ and where the pseudogap turns out to be minimal), at zero 
frequency (about which the relevant range of the pseudo-gap phenomena is centered),
and at a temperature close to $T^*$ (where the system recovers a Fermi-liquid 
behavior). An exact selection of $T$ is, in practice, not
required since $\Sigma_0$ turns out to depend rather weakly on
$T$ in the intermediate- to weak-coupling regime.
In this sense, we interpret $\Sigma_0$ as {\em a kind of Hartree shift\/},
even though for our choice of the potential the true Hartree shift vanishes 
identically.
We have consistently evaluated  the constant shift $\Sigma_0$ at 
$|{\bf k}|=k_{\mu'},\omega=0,$ and at the {\em same temperature\/}  where the 
self-energy $\Sigma({\bf k},\omega)$ of Eq.(1) is calculated\cite{infinity}.

The inclusion of the above constant shift $\Sigma_0$ stems from the need
of improving the single-particle Green's functions entering the construction
of the self-energy (1) when approaching $T_c$, only close to which pseudogap
phenomena become appreciable in the intermediate- to weak-coupling regime.
The choice of the self-energy (1) takes, in fact, into account fluctuation 
corrections only at the lowest order, a procedure which is certainly not 
completely satisfactory when approaching the critical temperature where all 
sort of fluctuations corrections become important.
To approach $T_c$, one may 
try to improve the self-energy (1) by dressing the single-particle 
Green's functions therein with a constant self-energy insertion appropriate to the 
non-critical (temperature) region. 
On the other hand, the inclusion of the full self-consistent Green's function
(without vertex corrections, however) leads to an overall depression of pseudogap phenomena
and is not theoretically justified \cite{Kyung,Pieri}.
From a pragmatic point of view, we have verified that 
in the intermediate- to weak-coupling regime the pseudogap would open at 
negative frequencies (and not at $\omega=0$, as expected from a simple physical
intuition), if the constant self-energy shift $\Sigma_0$ were not
properly included. The pseudogap opening at negative frequencies would, in turn, 
be in contrast with Monte Carlo results and experimental findings.
        
The characteristic behavior of the imaginary and real parts of the
self-energy at $|{\bf k}|=k_{\mu'}$ are shown in Figs.11 and 12, 
respectively, at different temperatures (for a given coupling).
Note that the convexity of the curves 
$\mbox{Im}\Sigma(k_{\mu'},\omega)$ about $\omega =0$ is inverted with respect 
to the Fermi-liquid 
behavior, implying strong deviations from Fermi-liquid behavior also at  
moderate values of the coupling (i.e., such that a bound-state in the two-body 
problem is not yet present). We have verified, however, that the Fermi 
liquid behavior is consistently recovered when the coupling is progressively 
decreased.

In the weak-coupling limit and for temperature close to $T_c$,
an analytic approximation
for the imaginary part of the self-energy can be obtained by inserting into 
Eq.(5) the weak-coupling 
expression of the pair-fluctuation propagator given by Eq.(12). 
At zero frequency and at the Fermi wave vector,
the imaginary part of the self-energy acquires then 
the following expression in the limit $T\rightarrow T_c$:
\begin{equation}
\label{imselfweak}
\mbox{Im} \Sigma ({\bf k_F},\omega=0)=
\frac{6\pi^3}{7\zeta (3)}\left ( \frac{T_c}{\epsilon_F} \right )^2
\frac{T_c}{2} \, \mbox{ln}\left ( \frac{T-T_c}{T_c} \right )
\end{equation}
which {\em diverges} upon approaching $T_c$ with a slow logarithmic rate. 
An expression analogous to (14) is also obtained at finite frequency 
(such that $\mid \omega \mid \ll \epsilon_F$) and $T=T_c$, with the replacement
of $\ln ((T-T_c)/T_c)$ by $\ln (\mid \omega \mid /\omega_c)$,
where $\omega_c \ll \epsilon_F$ is a suitable cutoff frequency.

To test the validity of the above analytic approximations, we may consider, e.g., 
the case of Fig.11 for $T/T_c=1.001$ and obtain from Eq.(14) the value 
$\mbox{Im} \Sigma / \epsilon_F\simeq -0.92$ for $\omega=0$.
This estimate is indeed in good agreement with the numerical result reported
in Fig.11 (cf. the full curve therein), for which 
$\mbox{Im} \Sigma / \epsilon_F\simeq -1$.
A fine tuning of the temperature very close to $T_c$ is, however, necessary to get a 
sizable increase of $\mid \mbox{Im} \Sigma \mid$ due to the logarithmic divergence
in Eq.(14). For instance, to double the above value a temperature 
$(T-T_c)/T_c=10^{-6}$ has to be reached. In the 3D model here considered,
the divergence of $\mbox{Im} \Sigma$ is therefore not numerically detectable for
all practical purposes. In addition, to test the validity of the counterpart of Eq.(14)
extended to finite frequency as explained above, we may consider the case of Fig.11 for
$T/T_c=1.001$ and two different frequencies, say, $\omega_1/\epsilon_F=0.075$
and $\omega_2/\epsilon_F=0.037$. In this case, we obtain from our analytic approximation
the value $(\mbox{Im}\Sigma(\omega_1)-\mbox{Im}\Sigma(\omega_2))/\epsilon_F=0.093$, 
which is rather close to the numerical result 0.106 as obtained from Fig.11.

The analytic approximation (14) [as well as its counterpart at $T=T_c$ and finite $\omega$]
need to be compared with the analytic form of $\Sigma({\bf k},\omega)$ 
obtained in the weak-coupling limit by Ref.\onlinecite{schmid} within the same non-self-consistent
t-matrix approximation adopted in the present paper.
According to Ref.\onlinecite{schmid}, the diffusive form (12) of the 
pair-fluctuation propagator would yield
\begin{equation}
\Sigma({\bf k},\omega)= \frac{\Delta_{pg}^2}{\omega+\xi({\bf k})+ i \gamma}
\label{sch}
\end{equation}
where $\Delta_{pg}$ is a parameter that depends on a wave-vector cutoff and 
$\gamma \propto (T-T_c)$.
This expression evidently does not reduce to Eq.(14) for $\omega=0$ and $T\rightarrow T_c$,
nor to the counterpart of Eq.(14) for $T=T_c$ and finite $\omega$.
A few comments to clarify the origin of these discrepancies are then in order.

The expression (\ref{sch}) has been derived more recently in Ref.\onlinecite{NormRand},
where it was also extensively used to fit ARPES data for $Bi$-based cuprates.
According to Ref.\onlinecite{NormRand},
Eq.(\ref{sch}) results by manipulating directly the expression (1) for the self-energy 
in Matsubara frequency, whereby the finite value $\pi T$ of the smallest (fermionic) Matsubara
frequency is exploited to make approximations on the ${\bf q}$-dependence of the
integrand. Analytic continuation to the real frequency axis is then performed on the
approximate result, yielding eventually the expression (\ref{sch}) above. 
This procedure is, however,
questionable, insofar as the very variable to be analytically continued is used to set restrictions
on the approximate form of the function (in this case, the ${\bf q}$-dependence of the
integrand). In our procedure, on the other hand, analytic continuation is performed 
{\em at the outset} [cf. Eq.(5)] and the relevant (controlled) approximations to get
the approximate result (14) are introduced only afterwards.

Note, in addition, that at $\mid {\bf k}\mid=k_F$ 
the expression (\ref{sch}) produces two peaks symmetrically located about
$\omega=0$. This expression cannot, therefore, be used to fit the
curves of $A(k_{\mu'},\omega)$ for the coupling values we are 
considering [cf. Figs.~13 and 16 below], whereby the symmetry of the two peaks is recovered
only in the extreme weak-coupling limit. In the analysis reported in Ref.\onlinecite{NormRand}, 
on the other hand, the experimental data are artificially symmetrized
and the expression (\ref{sch}) (together with an additional scattering rate $i\Gamma_1$)
is used to fit the ARPES data. We shall propose below an alternative phenomenological fit
to the curves of $A(k_{\mu'},\omega)$, which is suggested by our numerical calculations.

In our numerical calculations we have found that, at low enough temperature, there 
are three intersections
of the curves ($\mbox{Re}\Sigma (k_{\mu'},\omega)-\Sigma_0)$ with the
straight line $\omega$ (not shown in Fig.~12), with the two outer intersections 
giving rise to the two peaks of $A(k_{\mu'},\omega)$ (see Fig.13) while the 
central 
intersection corresponds to a strong suppression of 
$A(k_{\mu'},\omega)$ owing to the associated large value of
$\mbox{Im}\Sigma(k_{\mu'},\omega)$. By increasing 
temperature, on the other hand, only one intersection remains (resulting in 
only one visible peak in $A(k_{\mu'},\omega)$ - see Fig.13).
 
\begin{figure}
\label{selfi}
\centerline{\psfig{figure=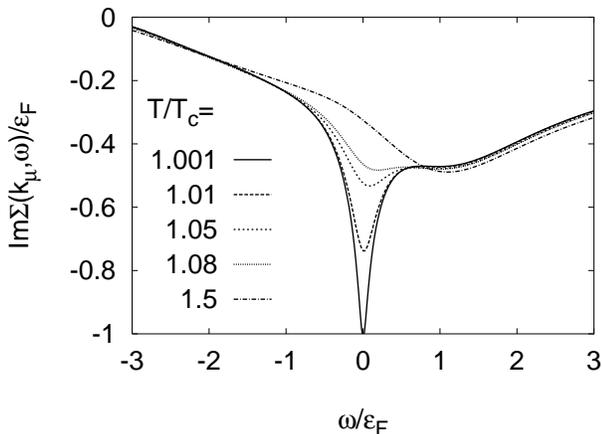,width=6cm,angle=-90}}
\caption{Imaginary part of the self-energy at $|{\bf k}|=k_{\mu'}$
as a function of frequency (in units of $\epsilon_F$) at different temperatures 
when $(k_Fa_F)^{-1}=-0.45$ ($T_c/\epsilon_F=0.23$). (Intermediate- to weak-coupling 
regime.)}
\end{figure}

\begin{figure}
\label{selfr}
\centerline{\psfig{figure=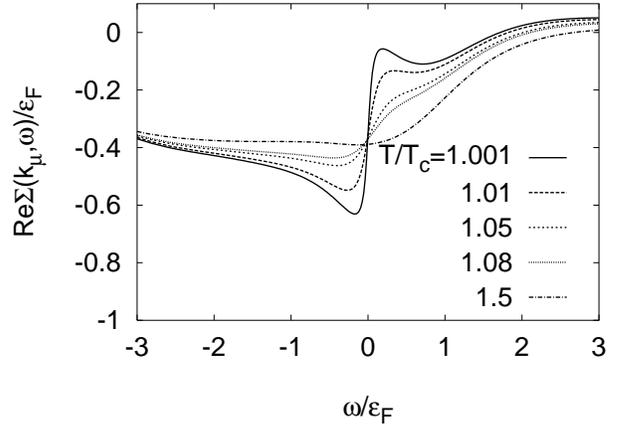,width=6cm,angle=-90}}
\caption{Real part of the self-energy at $|{\bf k}|=k_{\mu'}$
as a function of frequency (in units of $\epsilon_F$) at different temperatures when
 $(k_Fa_F)^{-1}=-0.45$ ($T_c/\epsilon_F=0.23$). 
(Intermediate- to weak-coupling regime.)}
\end{figure}
The associated spectral function at $|{\bf k}|=k_{\mu'}$
is reported in Fig.13 for the same temperatures and coupling of Figs.~11 and 
12.
\begin{figure}
\label{akwic}
\centerline{\psfig{figure=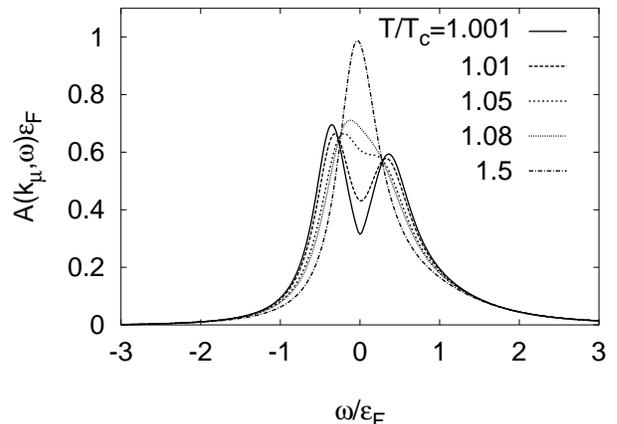,width=6cm,angle=-90}}
\caption{Spectral function at $|{\bf k}|=k_{\mu'}$
as a function of frequency (in units of $\epsilon_F$) at different temperatures. 
In this case, with $(k_Fa_F)^{-1}=-0.45$ 
($T_c/\epsilon_F=0.23)$. (Intermediate- to weak-coupling regime.)}
\end{figure}
The spectral function obtained in the intermediate-coupling regime
shows a well-developed two-peak structure near $T_c$ with a minimum at
zero frequency; yet the spectral weight distribution remains slightly
{\em asymmetric\/} about zero frequency, even when approaching the
critical temperature. At zero frequency the spectral
function has a sizeable finite value, 
indicating that no real gap opens at the Fermi surface. 
Note from Fig.13 that, upon increasing the temperature,
{\em the pseudogap fills in and closes at the same time\/}, with the two peaks 
of the spectral function merging in just one peak at a
crossover temperature $T_1^* \simeq T_0^*$ 
(which in this particular case is between $1.05 T_c$ and  $1.08 T_c$).
It is thus apparent that a breakdown of the normal-state Fermi liquid occurs 
well before the system is in the preformed-pair limit.
From the two-peak structure of $A(k_{\mu'},\omega)$
in the intermediate- to weak-coupling regime, 
a {\em pseudogap\/} $\Delta_{pg}$ could be empirically defined either  
as {\em half\/} the frequency separation between the maxima of the peaks, 
or as the separation of the maximum of the lower 
peak (at negative frequencies) from zero frequency. These two definitions 
coincide in the weak-coupling limit but slightly differ in the 
strong-coupling limit (see also Table II below). Throughout this paper we will 
adopt the second definition, which is the most relevant for comparison with 
photoemission experiments, accessing only negative frequencies.   

In the intermediate-coupling regime, when the chemical potential lies
inside the fermion band and the Fermi surface is well defined,
the wave-vector dependence of the spectral function shows a
strong asymmetry about the wave vector $k_{\mu'}$.
In Fig.14 the spectral function is reported as a function of frequency 
for different wave vectors about $k_{\mu'}$ at $T=T_c$.
\begin{figure}
\label{akwdiffk}
\centerline{\psfig{figure=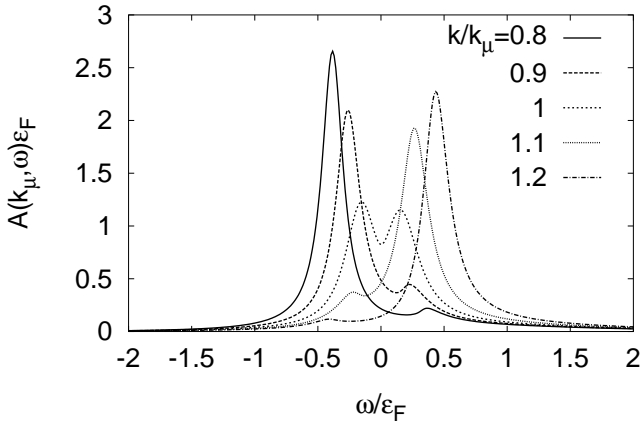,width=6cm,angle=-90}}
\caption{Spectral function at different wave vectors 
$|{\bf k}|$ about $k_{\mu'}$
as a function of frequency (in units of $\epsilon_F$), for $(k_Fa_F)^{-1}=-0.72$
and $T/T_c=1.001$. (Intermediate- to weak-coupling regime.)}
\end{figure}
It is clear from this figure that for $|{\bf k}|<k_{\mu'}$ a well-defined peak is
found at negative frequencies,
and that increasing the wave vector to $|{\bf k}|>k_{\mu'}$ 
this peak becomes a small and broad incoherent peak.
Thus, for $|{\bf k}|< k_{\mu'} $ 
the spectral weight of the coherent peak at negative frequencies
decreases as the wave vector
${\bf k}$ approaches $k_{\mu'}$,
while at the same time the spectral weight of the associated incoherent peak
located at positive frequencies increases, with
a transfer of spectral weight from negative to positive 
frequencies upon crossing the ``Fermi surface'' (which is defined as
the locus of minimum pseudogap, and almost coincides with the sphere 
$k=k_{\mu'}$; note that for the coupling value of Fig.~14, $k_{\mu'}$ is about 
10 \% smaller than $k_F$).
This clearly shows that the interaction of the fermions
with pair fluctuations gets increasingly stronger upon approaching
the ``Fermi surface'', so that {\em deviations from the Fermi liquid\/}
picture appear to be stronger at low energy.

In Fig.15 the positions of the two peaks of the spectral function 
are reported for different wave vectors about $k_{\mu'}$ at 
$T=T_c$.
The results of our non-self-consistent t-matrix approximation (squares and 
asterisks) are here compared with the BCS-like 
dispersion $\omega=\pm\sqrt{\xi({\bf k})^2+\Delta_{pg}^2}$ (continuous and 
dotted lines), 
where the BCS gap has been replaced by the pseudogap $\Delta_{pg}$ 
at $k_{\mu'}$. 
It is rather remarkable that the coherent peak 
of the spectral function at $|{\bf k}|<k_{\mu'}$
gets reflected into the incoherent peak at $|{\bf k}|>k_{\mu'}$
as the wave vector crosses the ``Fermi surface'' (with the characteristic
behaviour of an {\em avoided level crossing}), 
in such a way that the position of the peak at 
negative frequencies follows almost exactly the BCS-like 
dispersion, provided the value of the pseudogap is inserted 
as explained above. 
A similar behavior is displayed by the peak at positive energy.
This result supports the ``heuristic description'' 
of the pseudogap phase in terms of an effective BCS approach, that 
we shall discuss in the last Section.

\begin{figure}
\label{ekdiffk}
\centerline{\psfig{figure=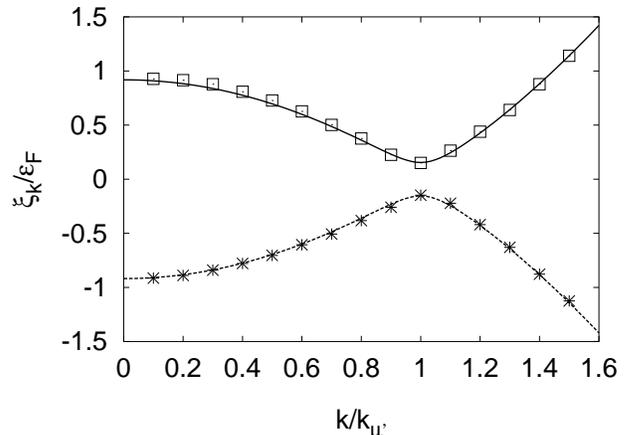,width=6cm,angle=-90}}
\caption{Peak positions of the spectral function 
at negative (asterisks) and positive (squares) frequencies versus  
wave vector for $(k_Fa_F)^{-1}=-0.72$ and $T/T_c=1.001$. Full and dotted lines represent the 
BCS-like fit. (Intermediate- to weak-coupling regime.)}
\end{figure}

To fit the prominent features of $A(k_{\mu'},\omega)$ with a simple analytic 
expression (from which the corresponding form of $\Sigma(k_{\mu'},\omega)$ replacing
Eq.~(15) could be extracted), we may consider two Lorentians of width $\gamma_L$ 
and $\gamma_R$, centered at $-\Delta_L$ and $\Delta_R$, and with weights $p_L$ and
$p_R$ (such that $p_L+p_R=1$), with the labels $L$ and $R$ referring to the 
left and right peaks of $A(k_{\mu'},\omega)$, in the order. 
In Tables I and II we 
report the values of the fitting parameters $\Delta_L$,$\Delta_R$,$\gamma_L$,
$\gamma_R$, and $\alpha=1-2 p_L$ for the curves of Fig.~13 
(fixed coupling and 
varying temperature) and of Fig.~16 (fixed temperature and varying coupling), respectively.

Note that the asymmetry of the two Lorentians (which is controlled by the 
parameters $\alpha$ and $\gamma_L/\gamma_R$) is considerable, increasing for 
increasing temperature or coupling (but for the last value of Table II).
For temperatures and couplings larger than those reported in the tables, 
however, 
the fit of $A(k_{\mu'},\omega)$ with two Lorentians become inadequate.  

\begin{table}
\begin{tabular}{|c|c|c|c|c|c|} 
$T/T_c$&$\Delta_L$&$\Delta_R$&$\gamma_L$&$\gamma_R$&$\alpha$\\ \hline
1.001&0.31&0.31&0.16&0.29&0.22\\ \hline
1.01&0.28&0.28&0.17&0.32&0.28\\ \hline
1.05&0.21&0.21&0.18&0.33&0.33\\ \hline
1.08&0.18&0.18&0.18&0.34&0.36  
\end{tabular}
\caption{Fitting parameters for the curves of Fig.~13. Energy variables are 
in units of $\epsilon_F$.}
\end{table}

\begin{table}
\begin{tabular}{|c|c|c|c|c|c|} 
$(k_F a_F)^{-1}$&$\Delta_L$&$\Delta_R$&$\gamma_L$&$\gamma_R$&$\alpha$\\ \hline
-1.1&0.035&0.035&0.042&0.047&0.060\\ \hline
-0.72&0.14&0.14&0.11&0.15&0.17\\ \hline
-0.23&0.47&0.61&0.22&0.33&0.15\\ \hline
0&0.78&0.84&0.25&0.18&0.008  
\end{tabular}
\caption{Fitting parameters for the curves of Fig.~16. Energy variables are 
in units of $\epsilon_F$.}
\end{table}

Note also that in most cases $\Delta_L=\Delta_R=\Delta_{pg}$.
In these cases a relatively simple form for $\Sigma(k_{\mu'},\omega)$ can 
be extracted, yielding:
\begin{eqnarray}
& &\Sigma(k_{\mu'},\omega)= -i (\gamma -\alpha\delta) + \alpha \Delta_{pg}
\nonumber\\
&+&\frac{(1-\alpha^2)(\Delta_{pg}^2-\delta^2) -2 i \Delta_{pg}\delta(1+\alpha^2)}
{\omega+\alpha\Delta_{pg}+ i(\gamma+\alpha\delta)}
\label{sigmafit}
\end{eqnarray}
where $\gamma=(\gamma_R+\gamma_L)/2$ and $\delta=(\gamma_R-\gamma_L)/2$.
Note that, even in the symmetric case with $\alpha=0$ and $\delta=0$, the 
expression\rf{sigmafit} does not reduce to the form (15) (due to the presence 
of an
extra term $-i\gamma$ in Eq.\rf{sigmafit}), unless $\Delta_{pg}\gg \gamma$ (this
condition would be consistent with the assumptions under which Eq.\rf{sigmafit} 
has been derived only when $T$ approaches $T_c$ \cite{schmid}). However, the condition
$\Delta_{pg}\gg \gamma$ is never satisfied by our fits, where $\Delta_{pg}$ and 
$\gamma$ are of the same order.

In Fig.16 the spectral function at $|{\bf k}|=k_{\mu'}$
is reported for different values of
$(k_Fa_F)^{-1}$ from intermediate to weak coupling,
slightly above the critical temperature. 
Note that, in the weak-coupling regime,
the spectral function acquires an almost
{\em symmetric two-peak structure\/}, which differs from the standard BCS 
result at $T=0$ essentially for the broadening of the peaks due to the finite 
lifetime of the pairs.
Note also that the pseudogap near the critical temperature 
decreases with coupling. 

An analysis of the pseudogap opening within a 2D attractive Hubbard model
in the weak-coupling regime has recently been reported in Ref.\onlinecite{Metzner}, 
by means of the non-self-consistent T-matrix approximation formulated on the real 
frequency axis.
The frequency dependence of the spectral function obtained in that paper
(at quarter filling) resembles the results of our Fig.16.

Finally, a comparison of the pseudogap $\Delta_{pg}$ at $T_c$ with
the BCS gap $\Delta_{{\rm BCS}}$ at $T=0$ and with the two-body gap 
$\epsilon_0/2=\epsilon_F/(k_Fa_F)^2$ (which is non-vanishing only for $a_F>0$)
is shown in Fig.~17 for all coupling regimes 
[when $\mu<0$, $\Delta_{BCS}(T=0)$ is set equal to $(\mu^2+\Delta(T=0)^2)^{1/2}$].
Note that in the weak-coupling limit $\Delta_{pg}(T=T_c)\ll 
\Delta_{BCS} (T=0)$, while in the intermediate-coupling regime  
$\Delta_{pg}(T=T_c) \simeq \Delta_{BCS}(T=0)$. 
Moreover, in the intermediate- to strong-coupling regime (where $a_F>0$),
both $\Delta_{BCS}$ and $\Delta_{pg}$ approach $\epsilon_0/2$ 
from {\em above} as the coupling is increased. 
Many-body effects thus increase the pair-breaking energy scale with
respect to the two-body limit. This result resembles the pair-size-shrinking 
effect noticed in Ref.\onlinecite{Andrepjb} at the mean-field level.

In this context,
it is interesting to mention that, taking $\epsilon_F\simeq 400$ $meV$
as a representative value for cuprate superconductors, the range 
$\Delta_{pg}\simeq 20-120$ $meV$ 
characteristic of cuprate superconductors corresponds to 
$0.05\lesssim\Delta_{pg}/\epsilon_F\lesssim 0.3$, which (as seen from
Fig.17) lies within the range identified in Fig.3 for cuprates.
 
\begin{figure}
\label{akwvc}
\centerline{\psfig{figure=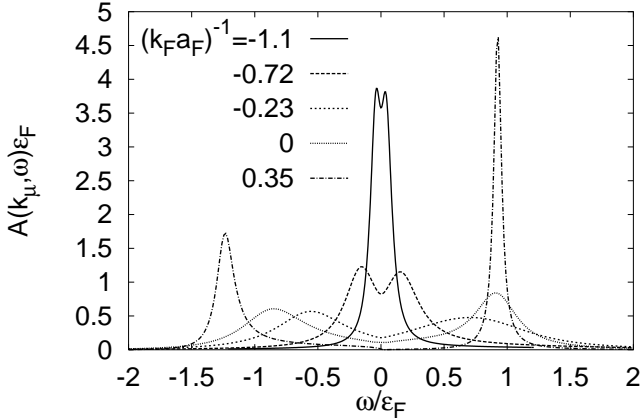,width=6cm,angle=-90}}
\caption{Spectral function at $|{\bf k}| =k_{\mu'}$ as a function of 
frequency $\omega$ (in units of $\epsilon_F$) for different values of 
$(k_Fa_F)^{-1}$ and $T/T_c=1.001$. 
(Intermediate- to weak-coupling regime.)}
\end{figure}
 
\begin{figure}
\label{psg}
\centerline{\psfig{figure=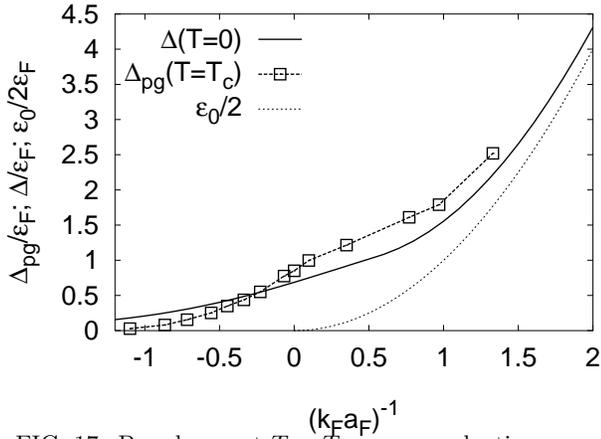,width=6cm,angle=-90}}
\caption{Pseudogap at $T=T_c$,
superconducting gap 
evaluated within the BCS approach at $T=0$ and gap in the strong-coupling limit, 
as functions of $(k_Fa_F)^{-1}$.} 
\end{figure}

\subsection{Criterion to distinguish weak from strong coupling}

The above systematic study of the single-particle spectral function from
weak to strong coupling suggests the following {\em criterion to 
distinguish by ARPES experiments whether 
a fermion system with an attractive interaction lies in the
strong- or weak-coupling regime\/}. 
This criterion rests on the analysis of the spectral function
at negative frequencies (just as determined by ARPES experiments) for
different values of the wave vector ${\bf k}$, and (as discussed in the next 
Section) it is meant to be useful for interpreting the experimental data for cuprates 
in conjunction with the two-gap model mentioned in the Introduction.

Consider first a system in the intermediate- to weak-coupling 
regime, for temperatures between $T^*$ and $T_c$,
i.e., within the pseudogap region.
In this case, the chemical potential lies inside the single-particle band and 
almost concides with the Fermi energy. For wave vectors smaller 
than $k_{\mu'}$, 
the spectral function has a quasi-particle peak with 
large spectral weight at negative frequencies and a smaller incoherent peak
at positive frequencies. 
Upon moving the wave vector across the ``Fermi surface''
($| {\bf k}| >k_{\mu'}$),
the quasi-particle peak shifts toward positive frequencies, while the 
incoherent peak is now present at negative frequencies (cf.~Fig.14)
and can accordingly be measured by ARPES. Restricting to negative frequencies
and realizing a cut in wave vector space 
which probes the main and the reflected (shadow) bands, 
starting from $| {\bf k}|<k_{\mu'}$ 
ARPES should initially find a well-defined 
quasi-particle peak which, upon increasing the wave vector to 
$| {\bf k}| >k_{\mu'}$, 
should be reflected as a small 
and broad incoherent peak. Moreover, at
$| {\bf k}| =k_{\mu'}$ the spectral weight at zero 
frequency remains a sizeable fraction of the peak maximum.
 
Consider then a system in the intermediate- to strong-coupling 
regime (when $\mu <0$), 
for temperatures between $T_0^*$ and $T_c$. In this case,
the chemical potential lies outside the single-particle band. 
For {\em any\/} wave vector, the spectral function has now a quasi-particle 
peak 
with large spectral weight at positive frequencies 
and a weaker incoherent peak at negative frequencies. 
For this reason, no appreciable difference in the shape of the
spectral function should be detected by varying the wave vector. 
Thus, starting, e.g., from ${\bf k}\equiv (k_x<0, 0, 0)$ ARPES should
find a broad incoherent peak
which, upon increasing the wave vectors to $(k_x>0, 0, 0)$, 
should not change appreciably.
In addition, the spectral weight vanishes or is much less than the
maximum of the incoherent peak in a range of frequencies
of the order of the pseudogap (cf.~Fig.9).

By this token, it is clear that,
for a fermionic system with an attractive interaction, 
the wave-vector dependence and the line-shape 
of the spectral function at negative frequencies 
have well-pronounced qualitative differences
depending on the coupling strength, differences which may be detected by
a detailed ARPES analysis of the spectral function, as discussed next.
Recall, however, that comparison of our results with ARPES data relies essentially
on the two-gap model mentioned in the Introduction, 
and can be complicated by the presence of additional sources of quasiparticle scattering 
in cuprates as well as by the fact that the continuum model relates strong
coupling to low density. Yet, our analysis can be useful to
understand the evolution of the spectral properties along the Fermi surface.

\section{Comparison with ARPES spectral function}
\label{conc}

The {\em systematic theoretical analysis of the spectral function\/} 
from weak to strong coupling presented in this paper can be used to analyze
the spectral intensities measured by ARPES 
in Bi-based superconducting cuprates, for which a 
systematic experimental analysis is also available.

In particular, we consider ARPES intensities measured in Bi2212
near the $M$ points of the Brillouin zone
as well as along the Fermi surface, moving from the $M$ points toward the
$N$ (nodal) points, in different doping regimes and at different temperatures. 
According to our understanding of the phenomenology of cuprate
superconductors, the effective coupling between fermions should increase 
from the weak- to strong-coupling regime, when the doping
is reduced from over to optimum doping and eventually to underdoping.
Moreover, as discussed in the Introduction, when moving from $N$ toward 
$M$ points along the Fermi surface, 
a continuous crossover from weakly to strongly 
coupled fermionic states should be observed even at fixed doping.

We summarize as follows the main results extracted from our systematic
work in the different coupling regimes, 
which can be compared with ARPES experiments 
performed in Bi2212 materials.

\begin{center}
{\em Strong- to intermediate-coupling regime (about M points)\/}:
\end{center}

In the strong- to intermediate-coupling regime, where the chemical
potential is below the bottom of the single-particle
band ($\mu<0$), our results
show that the spectral function is strongly asymmetric in frequency, with
two peaks, one completely incoherent at negative frequencies and 
the other one coherent at positive frequencies. 
In this case, the wave vectors are meant to be reckoned with respect to 
(one of) the M points.
In this regime, the prominent features
to be compared with experiments are:

({\em i}) The line shape of the spectral function 
at negative frequencies
is quite broad, and the height of the incoherent peak
noticeably decreases with increasing temperature (see~Fig.8)
or increasing coupling (see~Fig.9).
These features are in qualitative agreement with 
the behavior of the spectral intensity observed
by ARPES experiments in the pseudogap phase of {\em underdoped\/} cuprates, by
decreasing doping and increasing temperature.
Several ARPES measurements show, in fact, that the height of the peak
in the spectral intensities collected about the $M$ points
decreases with underdoping, with heavily underdoped cuprates
displaying a very broad structure with no detectable peak 
(cf., e.g., Fig.2 (left panel) of Ref.\onlinecite{Vobornik} 
and Fig.1 (panel $a$) of Ref.\onlinecite{Ding}, for the
doping dependence of the spectral weight about the $M$ points).
ARPES measurements for the temperature dependence of the 
(quite broad) spectral intensities about the $M$ points  
further indicate that the spectra are (slightly) suppressed for increasing 
temperature (cf.~Fig.2 (panel $b$) of Ref.\onlinecite{Ding}).

({\em ii}) The spectral weight near zero frequency is strongly suppressed
and a real gap opens in the spectral function
in the strong-coupling regime (see~Figs.8 and 9).
Experimental evidence for a strong suppression of the spectral
weight near zero frequency can indeed be found, e.g.,
in Fig.2 (left panel) of Ref.\onlinecite{Vobornik} for
(heavily) underdoped samples with $T_c=56 K$.

\begin{center}
{\em Intermediate- to weak-coupling regime (between M and N points)\/}:
\end{center}

In the intermediate- to weak-coupling regime, the chemical potential
lies within the single-particle band $(\mu >0)$ and the wave vectors
are referred to the center of the Brillouin zone. 
In this case, the salient features
of our calculations to be compared with experiments are:

({\em i}) A single quasi-particle peak is present in the spectral function
above the crossover temperature $T^*$ (see~Fig.13), 
implying a well-defined Fermi surface. 
Under-, optimally, and over- doped cuprates for wave vectors near
the nodal points unquestionably
display quasi-particle peaks in the ARPES spectral
intensities (cf., e.g., Fig.1 of Ref.\onlinecite{Norm1}).

({\em ii}) Approaching the critical temperature from above, the 
interaction between fermions and (damped) pair fluctuations determines
a suppression of spectral weight near zero frequency and 
therefore the opening of a pseudogap, characterized by a finite spectral
weight at zero frequency (see~Fig.13). 
In addition, the quasi-particle peak disperses as a function of the
wave vector and, as the wave vector moves across the Fermi surface,
is reflected as an incoherent broad peak (see~Fig.14).
ARPES spectral intensities in {\em underdoped\/} cuprates, measured 
about the $N$ points for temperatures
between $T^*$ and $T_c$, display this feature, even though the 
reflection cannot be accurately identified (probably owing to the low
spectral weight of the incoherent peak).
In particular, a spectral weight suppression at low frequencies 
and a finite spectral weight at zero frequency has been found 
by several ARPES measurements (cf., e.g., Fig.1 (panel $b$) 
and Fig.3 (panel $a$) of Ref.\onlinecite{NormRand}). 
Experimental evidence for the reflection of the quasi-particle
peak into an incoherent peak has also been found by ARPES measurements of 
the peak along the $MY$ direction in the pseudogap phase of
slightly underdoped cuprates (cf.~Fig.2 (panel $b$) of Ref.\onlinecite{Camp1}),
for which the intermediate- to weak-coupling regime should apply.
 
({\em iii}) Increasing the coupling from the weak- to the
intermediate-coupling regime, the pseudogap evaluated  at $T_c$
increases and the ratio between the pseudogap at $T_c$ and the
BCS gap evaluated at $T=0$ also increases (see Fig.~17), about coinciding in
the intermediate-to-strong coupling region.
In all underdoped cuprates, and for any wave vector, the 
experimentally determined pseudogap at
$T_c$ clearly increases with decreasing doping, and in
heavily underdoped cuprates it almost coincides with the 
superconducting gap measured at zero temperature
(cf., e.g., Fig.3 (panel $b$) of Ref.\onlinecite{Ding}).

\section{Discussion and conclusions}
\label{dis}

In this paper,
the evolution (from superconducting fluctuations to the bosonic limit)
of the pseudogap opening and the spectral function
has been studied in a {\em systematic way\/}. 
A system of fermions in a three-dimensional continuum, mutually
interacting via an attractive contact potential, has been examined.
In this way, the numerical calculation of the single-particle
Green's function has been considerably simplified,
yet preserving the main physical effects underlying the pseudogap
opening.
The pair-fluctuation propagator,
the (one-loop) self-energy, and the spectral function
have been evaluated as functions of coupling strength and temperature,
from weak to strong coupling, and 
analytic and numerical results have been presented.

In the {\em strong-coupling regime}, the pair-fluctuation propagator
has been shown to have bosonic character and the line-shape 
of the incoherent peak of the spectral function to be 
strongly asymmetric about its maximum, with its spectral weight 
decreasing by increasing coupling
(or decreasing density) and increasing temperature. 
In this regime, two crossover temperatures $T^*_1$ (at which
the two peaks in the spectral function merge in just one peak)
and $T^*_0$ (at which the maximum of the incoherent peak crosses zero
frequency) have been identified, with $T^*_1>T^*_0\gg T_c$ 
and with $T^*_0$ of the order of the binding energy of preformed pairs
(ARPES experiments, however, can only measure $T_0^*$).

In the {\em intermediate-coupling regime}, the line shape of the spectral
function about the ``Fermi surface'' resembles the 
line shape of the spectral intensity (which is, in turn, related
to the spectral function) measured by ARPES in underdoped 
cuprates between $T_c$ and $T^*$
for different wave vectors. In particular, we have reproduced the main features
characterizing the ARPES pseudogap, namely, a finite spectral intensity at zero
frequency and a finite pseudogap at $T=T_c$ which is of the same order 
of the superconducting gap at zero temperature.
We have also found that in the intermediate- to weak-coupling regime
pseudogap effects are present only in a narrow
temperature range above the critical temperature, a result related
with the 3D character of the pair fluctuations (in 2D this temperature
range should, in fact, be considerably wider). 
To obtain a wider temperature range for pseudogap
effects in 3D it is thus necessary to increase the coupling as to reach
the strong-coupling regime, at the price of destroying the Fermi surface.

In the {\em weak-coupling regime}, the pair fluctuation propagator 
acquires the diffusive
Ginzburg-Landau character and the line shape of
the spectral function gets progressively more symmetric
as the coupling is decreased.  
In this regime, the two crossover temperatures $T_1^*$ and $T_0^*$
coincide and are of the order of $T_c$, with 
the pseudogap closing and filling-in quickly as the temperature is increased 
above $T_c$.

It is thus clear that the pseudogap already occurs in the one-loop
approximation for the self-energy, namely, the non-self-consistent
t-matrix approximation which we have adopted in this paper. 
A similar non-self-consistent (as well as a self-consistent) calculation
for the spectral function has been reported in Ref.\onlinecite{Yamada}.
However, detailed comparison of our results with the results
of Ref.\onlinecite{Yamada} appears not to be possible,
since in Ref.\onlinecite{Yamada} the shift of the chemical potential was not
properly taken into account when evolving from the weak- to 
intermediate-coupling regimes.

Most significantly, the results presented in this paper,
concerning the temperature and wave-vector
dependence of the spectral function in the pseudogap phase,
are in qualitative agreement with 
Monte Carlo simulations of the 2D attractive ($s$-wave) Hubbard model.
In particular, in Refs.\onlinecite{Vilk} and \onlinecite{Kyung} 
the spectral function obtained by Monte Carlo simulations is reported in the 
intermediate-coupling regime for different temperatures and wave vectors.
Monte Carlo simulations clearly show that in the pseudogap phase
the spectral function has a two-peak structure, with the incoherent peak
smoothly emerging from the main peak as the temperature 
is lowered below $T^*$.
In addition, moving the wave vector across the Fermi surface, the main peak is
reflected in a shadow incoherent peak, as reported in Ref.\onlinecite{Singer}.
Monte Carlo simulations on the 2D attractive ($s$-wave) Hubbard model thus give
further support to our non-self-consistent t-matrix approximation,
suggesting that dimensionality and lattice effects do not modify 
appreciably the main qualitative features of the pseudogap phase, obtained by 
our work for a 3D continuum with a contact potential. 

The evolution of the spectral function
from weak to strong coupling in the pseudogap region 
can be rationalized in a {\em heuristic\/} way
by a BCS-like approach {\em at zero temperature\/}. 
The BCS spectral function is given by
$A({\bf k},\omega)=u({\bf k})^2\delta (\omega-E({\bf k}))+v({\bf k})^2\delta 
(\omega+E({\bf k}))$,
where $E({\bf k})=\sqrt{\xi({\bf k})^2+\Delta ^2}$
is the quasi-particle energy,
$u({\bf k})^2=\frac{1}{2}(1+\xi({\bf k})/E({\bf k}))$ and
$v({\bf k})^2=\frac{1}{2}(1-\xi({\bf k})/E({\bf k}))$ are the weights of the
quasi-particle poles at positive and negative frequencies, 
respectively. In Fig.18 we sketch this spectral function vs. 
frequency for three characteristic wave vectors, namely, 
$| {\bf k}| >>k_F$ (top panel),
$| {\bf k}| =k_F$ (central panel), and 
$| {\bf k}| <<k_F$ (bottom panel).
This behavior of the BCS spectral function for different wave vectors
is seen to reproduce the qualitative features of the spectral function
which we have obtained for the pseudogap phase {\em above\/} $T_c$ in the 
intermediate- and weak-coupling regimes (cf. Figs.13-16), 
{\em provided\/} the BCS gap 
$\Delta$ at $T=0$ is suitably
replaced by the pseudogap $\Delta_{pg}$
(obtained from the numerical calculation above $T_c$) {\em and\/} the Fermi 
energy is replaced by the renormalized chemical potential 
$\mu' (T,k_Fa_F)$, which is coupling and temperature dependent
(and also provided the delta-like peaks of Fig.18 are suitably broadened).
\begin{figure}
\label{p3}
\centerline{\psfig{figure=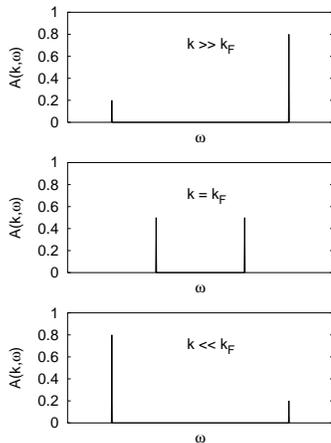,width=6cm,angle=-90}}
\caption{Schematic representation of the BCS spectral 
function as a function of frequency
for three different wave vectors: 
$| {\bf k} | >>k_F$ (top panel),
$| {\bf k}| =k_F $ (central panel), 
and $| {\bf k}| <<k_F$ (bottom panel).}
\end{figure}
Note further that even the spectral function in the
strong-coupling regime can be heuristically understood 
by comparing the BCS spectral
function of Fig.~18 for $| {\bf k}| \gg k_F$ 
with our results of Figs.8 and 9. 
The distribution
of the spectral weight and the position of the peaks in the spectral
function above $T_c$ appear thus to be in generic correspondence with the features 
of the BCS spectral function at $T=0$. The intimate 
reason why the BCS approach at $T=0$ is able to
give information on the pseudogap state above $T_c$ remains to be
clarified, for instance, by extending the present non-self-consistent t-matrix 
from above to below $T_c$.  

Other kinds of fluctuation propagators (like, 
charge-density wave \cite{Caprara,notaqcp}, spin-density wave \cite{Pines,Kampf,Schr},
and phase fluctuations above the Kosterlitz-Thouless 
transition \cite{Sharapov})
result into peak structures in the two-particle Green's function and into an
associated pseudogap opening in the single-particle spectral function.
In particular, the pioneering work by Kampf and Schrieffer \cite{Kampf} considering 
antiferromagnetic fluctuations coupled to fermions has shown that the associated spectral
function evolves from one peak in the Fermi-liquid regime 
to two peaks in the fluctuation regime.
In addition, the antisymmetric structure of the imaginary part of the
susceptibility used by Kampf and Schrieffer is reminescent of the
behavior of the imaginary part of our pair-fluctuation propagator
in the weak-coupling regime only. 

Further detailed ARPES (and, possibly, inverse photoemission) experiments
are awaited to ultimatly distinguish the microscopic origin of the pseudogap
in underdoped cuprates and to unambiguosly identify the
characteristic features of the spectral function obtained by our
analysis in different doping and coupling regimes.

\section*{Acknowledgements}
The authors are indebted to A. A. Varlamov for discussions. 
A. Perali gratefully acknowledges financial support 
from the Italian INFM under contract PAIS Crossover No. 269.


\begin{thebibliography}{99}
\bibitem  {nota2d3d} 3D fluctuations are weaker than 2D ones as $T\rightarrow T_c$.
	  However, if critical properties are not of interest, the strength
          of 3D fluctuations can be increased by increasing coupling.
\bibitem  {notakfxi} In terms of the phenomenological parameter 
	  $k_F\xi_{pair}$ ($\xi_{pair}$ being the average pair size 
	  at $T=0$), the value $(k_Fa_F)^{-1}\simeq 2$
	  corresponds to $k_F\xi_{pair}\simeq 0.4$, 
          which is close to the value $k_F\xi_{pair}=1/\pi$ separating
	  the crossover region from the Bose-Einstein 
	  (strong-coupling) regime at $T=0$, as discussed 
	  in Ref.\onlinecite{Pist94}.
\bibitem  {infinity} It can be readily verified that the analytic continuation
of the self-energy (1) vanishes when $|\omega|\to \infty$. For this reason, 
there is no {\em additional\/} constant shift to be added to $\Sigma_0$, when 
the real part of $\Sigma({\bf k},\omega)$ is calculated as the Kramers-Kronig 
transform of the imaginary part (5).
\end{thebibliography}
\end{document}